# Doped rare gas clusters up to completion of first solvation shell, $CO_2$-$(Rg)_n$, n = 3 – 17, Rg = Ar, Kr, Xe

A.J. Barclay,[1] A.R.W. McKellar,[2] and N. Moazzen-Ahmadi[1]

[1] *Department of Physics and Astronomy, University of Calgary, 2500 University Drive North West, Calgary, Alberta T2N 1N4, Canada*

[2]*National Research Council of Canada, Ottawa, Ontario K1A 0R6, Canada*

**Abstract**

Spectra of rare gas atom clusters containing a single carbon dioxide molecule are observed using a tunable mid-infrared (4.3 μm) source to probe a pulsed slit jet supersonic expansion. There are relatively few previous detailed experimental results on such clusters. The assigned clusters include $CO_2$-$Ar_n$ with n = 3, 4, 6, 9, 10, 11, 12, 15, and 17, as well as $CO_2$-$Kr_n$ and -$Xe_n$ with n = 3, 4, and 5. Each spectrum has (at least) partially resolved rotational structure, and each yields precise values for the shift of the $CO_2$ vibrational frequency ($\nu_3$) induced by the nearby rare gas atoms, together with one or more rotational constants. These results are compared with theoretical predictions. The more readily assigned $CO_2$-$Ar_n$ species tend to be those with symmetric structures, and $CO_2$-$Ar_{17}$ represents completion of a highly symmetric ($D_{5h}$) solvation shell. Those not assigned (e.g. n = 7, 13) are probably also present in the observed spectra, but with band structures which are not well-resolved and thus not recognizable. The spectra of $CO_2$-$Ar_9$, -$Ar_{15}$, and -$Ar_{17}$ suggest the presence of sequences involving very low frequency ($\approx$2 cm$^{-1}$) cluster vibrational modes, an interpretation which should be amenable to theoretical confirmation (or rejection).

## 1. Introduction

There has long been interest in weakly bound molecular clusters. Simple hypothetical clusters, such as those governed by a Lennard Jones 12-6 potential, or real clusters such as those composed solely of rare gas (Rg) atoms, turn out to have complicated energy landscapes with many structural isomers once their size reaches beyond n = 4 or 5.[1] Clusters which contain a single carbon dioxide molecule and multiple Rg atoms, $CO_2$-$Rg_n$ form a family which is amenable to study by infrared spectroscopy, with $CO_2$ acting as a powerful chromophore, particularly thanks to its strong $\nu_3$ (asymmetric stretch) vibrational transition. But spectroscopic studies of such doped Rg clusters have been rare for n > 2, with the exception of those containing helium whose superfluid properties make it a special case.[2,3]

The present paper reports new high resolution infrared spectra of $CO_2$-$Rg_n$ clusters (Rg = Ar, Kr, Xe) in the range of n = 3 to 17. We recently reported[4] brief results for the largest of these clusters, $CO_2$-$Ar_{15}$ and $CO_2$-$Ar_{17}$, which represent completion of the first solvation shell of Ar atoms around the $CO_2$ molecule. Here we expand on that study with results for $CO_2$-$Ar_n$ (n = 3, 4, 6, 9, 10, 11, 12), for $CO_2$-$Kr_n$ (n = 3, 4, 5 (two isomers)), and for $CO_2$-$Xe_n$ (n = 3, 4, 5). For each cluster, we obtain precise information on the shift of the $CO_2$ $\nu_3$ vibrational mode as well as structural information from cluster rotational constant(s). The observed structures are generally in good agreement with calculations of those expected on the basis of $CO_2$-Rg and Rg-Rg intermolecular potentials.

The starting point for a $CO_2$-$Rg_n$ cluster is the $CO_2$-Rg dimer. $CO_2$-Ar was originally studied in 1979 by means of its pure rotational microwave spectrum,[5] and found to be T-shaped with the Ar atom located on the 'equator' of the $CO_2$, adjacent to the C atom, at a distance of about 3.5 Å. There have been many further microwave[6,7] and infrared[8-14] studies of $CO_2$-Ar. Spectroscopy of $CO_2$-$Kr^{[6,8,15-17]}$ and $CO_2$-$Xe^{[15,18]}$ yields analogous structures with C-Rg distances of about 3.6 and 3.8 Å, respectively. For a trimer, $CO_2$-$Rg_2$, it is clear that the second Rg atom will occupy a position equivalent to that of the first with an Rg-Rg distance similar to that of the corresponding $Rg_2$ dimer. This enables the trimer to retain perfectly the structures of the respective $CO_2$-Rg and Rg-Rg dimers. Microwave[19] and infrared,[20,21] spectra of $CO_2$-$Ar_2$ have confirmed this structure, with an Ar-Ar distance of about 3.8 Å. Very recently, infrared spectra of $CO_2$-$Kr_2$ and -$Xe_2$ have also been reported by our group.[21] But once we get to a tetramer, $CO_2$-$Rg_3$, the preferred structure is no longer obvious, as we will see below.

## 2. Experimental details

Spectra recorded for this study use a pulsed supersonic jet expansion for generation of the clusters and an infrared radiation source to probe the jet. Typical gas expansion mixtures contain about 0.03% carbon dioxide plus 1% argon, krypton, or xenon in helium carrier gas with a backing pressure of about 20 atmospheres. The jet is produced by two side-by-side pulsed solenoid valves (General Valve Series 9) controlled by Iota One valve drivers. A multi-channel block is attached to the bottom face of each valve flange. Six narrow cylindrical channels in each block provide six holes located symmetrically about the center of the block with smallest diameter hole (0.5 mm) close to the center and the largest (1.5 mm) far from the center. A pair of 5 cm long adjustable jaws is mounted on each block. In previous work, we normally set the jaw separation at 25 μm, but here we use a smaller slit width (12.5 μm) to encourage formation of larger clusters. The slits are aligned parallel to the optical axis of the multi-pass optical cell. This cell (Aerodyne Research Inc., AMAC-100) consists of two astigmatic mirrors installed at the ends of a 70 cm long base plate. It normally provides 182 passes, with toroidal mirrors circulating the optical beam to fill the volume of the cell and creating a pattern of spots which evenly fill the mirror surfaces. However, in our system this pattern is modified in order to confine the probe area in the center of the cell to a rectangle with a better overlap between the infrared beam and the jet. The modified number of passes is estimated to be about 100. To prevent diffusion pump oil from depositing on the mirrors, they are partially shielded by glass tubes and kept at a temperature of 45° C using small heaters. An NW250 six-way cross houses the absorption cell and nozzles, and this chamber is mounted on a Varian VHS-10 diffusion pump with a pneumatic gate valve separating the pump and chamber. The diffusion pump is backed by an Edwards E2M275 mechanical pump located in a separate room at a distance of about 3 meters in order to minimize noise and vibration.

The infrared radiation source is a continuous-wave singly resonant optical parametric oscillator (Lockheed Martin Aculight Argos Model 2400-SF-10 CW), used here with Module D to generate tunable IR radiation in the 4.25 μm region. The idler output, with less than 1 MHz linewidth, wide tunability (2260-2580 cm$^{-1}$), and high output power (20-300 mW), allows recording of rotationally resolved spectra for smaller clusters, and partially resolved spectra for larger ones. Coarse idler frequency tuning is done manually by a fine pitch screw which linearly translates a fan-out PPLN crystal and by an intracavity etalon. Fine tuning and scanning of the idler frequency are achieved by applying a 0-200 V signal from an external PZT driver (Thorlabs

MDT693B) to the seed laser PZT element. The data acquisition is based on a rapid-scan signal averaging mode with continuous background subtraction. In a typical experiment, the probe wavelength is scanned over a range of about 30 GHz (1 cm$^{-1}$) by strain variation of the fiber length in the seed laser. The scan drive, with a rate of 100 Hz and (sinusoidal) amplitude of 40 V, is generated by a function generator (WaveTek 195) and amplified by the PZT driver.

The pulsed molecular beam is 4 ms in duration and its central portion is probed for 2 ms near the zero crossing of the applied sine wave. Three LN$_2$-cooled InSb detectors monitor the infrared signals probing the jet, a reference gas cell, and a passive etalon. The jet signal is digitally scaled and subtracted from the signal from an InGaAs detector monitoring the signal channel of the OPO. This quantum correlated twin-beam (idler and signal) noise subtraction allows for 13 dB reduction in power fluctuations of the absorption signal in the rapid-scan mode.[22] The signal from the reference gas cell containing room temperature CO$_2$ is used for absolute wavenumber calibration and that from the passive etalon for interpolation. The detector signals are digitized at a rate of 4 MHz using a 12-bit data acquisition card (DAS4010/12) mounted on a PC motherboard. Spectrum simulation and fitting are done with the Pgopher software package.[23]

### 3. Cluster calculations

A number of previous publications have reported predicted structures for CO$_2$-Ar$_n$ clusters.[20,24-28] The most recent of these, by Wang and Xie,[27,28] are based on the same CO$_2$-Ar potential that we use, so their equilibrium results are very similar to ours, with only small differences due to our new analytical fit of the CO$_2$-Ar surface and our use of a slightly different Ar-Ar potential (see below).

We carried out our own cluster calculations since we are not aware of previous ones for CO$_2$-Kr$_n$ or -Xe$_n$, and since those mentioned for CO$_2$-Ar$_n$ often do not include full details of parameters which are important for spectral simulations, such as rotational constants and dipole moment components. Some CO$_2$-Ne$_n$ calculations were also carried out for comparison purposes, though we have no experimental results for n > 2. For Rg-Rg interaction potentials, we used recent results from Dieters and Sadus.[29] For CO$_2$-Rg, we used our own analytical fits to *ab initio* calculations by Chen et al.[30] (CO$_2$-Ne), Cui et al.[31] (CO$_2$-Ar), and Chen et al.[32] (CO$_2$-Kr). In the case of CO$_2$-Xe, *ab initio* potential surfaces have been reported by Chen and Zhu,[33] and by Wang et al.,[34] but we were unable to obtain their detailed results on request, so in order to obtain approximate information for CO$_2$-Xe$_n$ clusters, we simply scaled the fitted CO$_2$-Kr potential surface by the ratios of the distance and energy of the calculated potential minima of CO$_2$-Xe and

-Kr (1.059 and 1.116, respectively). Our own analytic potential fits (based on the MLR model[35]) were used here because the cluster calculations work best with potential functions which remain well-behaved even for physically unrealistic geometries, which was not always the case for the analytic or interpolated functions in the original *ab initio* papers.

The calculations start with a single $CO_2$ surrounded by n Rg atoms in random positions, and evaluate the total cluster potential energy as the sum of all the molecule-molecule potential energies (here we consider Rg atoms to be molecules). The multidimensional structure is then adjusted "downhill" in energy using the Powell method[36] to find the "nearest" local energy minimum. By repeating the calculation with hundreds or thousands of different starting structures, the global minimum and many other local minima are located. The results ignore possible non-additive many-body effects, since only two-body energies are included. Furthermore, the results correspond to equilibrium structures, which do not include the effects of zero-point vibrational motions. Zero-point intermolecular bond lengths almost always increase relative to equilibrium values and therefore zero-point rotational constants decrease, since they are proportional to inverse moments of inertia. We can get an idea of the magnitude of such effects by looking at results for dimers. Thus for $CO_2$-Ar, -Kr, and -Xe, the experimental rotational constants $B$ and $C$ are about 0.969, 0.976, and 0.988 of those given by (equilibrium) theory, respectively. Similarly, for $Ar_2$ and $Kr_2$ the experimental $B$-values are 0.963 and 0.977 of theoretical equilibrium ones, respectively.[37,38] When we talk about scaling calculated rotational parameters below, it simply means multiplying them by a suitable factor, usually about 0.97. Note also that differing zero-point energies could even affect the energy ordering of different structural isomers in cases where their calculated equilibrium energies are similar.

The availability of *ab initio* surfaces for the excited ($CO_2$ $\nu_3$) vibrational state enables estimation of a cluster vibrational shift as the difference between the ground and excited state binding energy. This represents a small difference between large numbers, and (again) does not include zero point vibrational effects, so we should be cautious when comparing calculated and experimental vibrational shifts, as done in Sec. 5 below.

Results of our structural calculations are summarized in Tables 1 − 3, and illustrated below as the cluster spectra are presented. Point group symmetries of our calculated clusters fall into one of the following cases: $C_1$ (no symmetry elements), $C_s$ (a single plane of symmetry), $C_{2v}$ (one twofold rotational axis and two symmetry planes), or symmetric rotor ($C_{3v}$, $C_{5v}$, $D_{3h}$, or $D_{5h}$). No vibrational shifts are given in Table 3 for $CO_2$-$Xe_n$ since the $CO_2$-Xe potential was just scaled from that of $CO_2$-Kr. The normalized transition dipole moment components, $\mu_a$, $\mu_b$, and $\mu_c$, represent the cosine of the angle between the O-C-O axis and the $a$-, $b$-, or $c$-inertial axis of the cluster.

Table 1. Calculated equilibrium structures of $CO_2$-$Ar_n$ clusters.[a]

| n | Sym | Binding energy | Vibrational shift | $A$ | $B$ | $C$ | $\mu_a$ | $\mu_b$ | $\mu_c$ |
|---|---|---|---|---|---|---|---|---|---|
| 3 | $C_s$ | 814.8 | -1.27 | 0.0345 | 0.0271 | 0.0241 | 0.85 | 0.0 | 0.52 |
| 4 | $C_{2v}$ | 1177.4 | -1.71 | 0.0255 | 0.0173 | 0.0169 | 0.0 | 0.0 | 1.00 |
| 5 | $C_s$ | 1542.1 | -2.21 | 0.0190 | 0.0130 | 0.0100 | 0.0 | 0.94 | 0.33 |
| 6 | $C_s$ | 1904.5 | -2.54 | 0.01296 | 0.01104 | 0.00800 | 0.97 | 0.0 | 0.28 |
| 7 | $C_1$ | 2315.5 | -3.26 | 0.01024 | 0.00808 | 0.00643 | 0.30 | 0.87 | 0.39 |
| 8 | $C_s$ | 2762.3 | -3.56 | 0.00830 | 0.00684 | 0.00543 | 0.0 | 0.76 | 0.65 |
| 9 | $C_{3v}$ | 3267.4 | -3.29 | 0.00629 | 0.00629 | 0.00492 | 0.0 | 0.0 | 1.00 |
| 10 | $C_{2v}$ | 3734.9 | -4.03 | 0.00575 | 0.00479 | 0.00431 | 0.0 | 0.0 | 1.00 |
| 11 | $C_{5v}$ | 4277.8 | -5.32 | 0.00458 | 0.00458 | 0.00394 | 0.0 | 0.0 | 1.00 |
| 12 | $C_s$ | 4268.1 | -5.69 | 0.00412 | 0.00351 | 0.00337 | 0.72 | 0.0 | 0.70 |
| 13 | $C_s$ | 5055.4 | -6.55 | 0.00381 | 0.00303 | 0.00281 | 0.86 | 0.0 | 0.51 |
| 14 | $C_s$ | 5577.0 | -4.66 | 0.00347 | 0.00276 | 0.00253 | 0.99 | 0.0 | 0.15 |
| 15 | $D_{3h}$ | 6106.1 | -6.37 | 0.00314 | 0.00240 | 0.00240 | 1.00 | 0.0 | 0.0 |
| 16 | $C_s$ | 6477.9 | -7.51 | 0.00266 | 0.00228 | 0.00206 | 0.93 | 0.37 | 0.0 |
| 17 | $D_{5h}$ | 7062.7 | -10.17 | 0.00270 | 0.00189 | 0.00189 | 1.00 | 0.0 | 0.0 |
| 18 | $C_{2v}$ | 7494.4 | -10.28 | 0.00226 | 0.00188 | 0.00169 | 1.00 | 0.0 | 0.0 |

[a] Sym = point group symmetry. Binding energies and rotational constants $A$, $B$, $C$, are in units of cm$^{-1}$. Dipole moment is normalized to unity.

Table 2. Calculated equilibrium structures of $CO_2$-$Kr_n$ clusters.[a]

| n | Sym | Binding energy | Vibrational shift | $A$ | $B$ | $C$ | $\mu_a$ | $\mu_b$ | $\mu_c$ |
|---|-----|---------|-----------|------|------|------|------|------|------|
| 3 | $C_s$ | 1008.6 | -2.09 | 0.0173 | 0.0156 | 0.0115 | 0.98 | 0.0 | 0.21 |
| 4 | $C_{2v}$ | 1479.9 | -2.74 | 0.01371 | 0.00812 | 0.0746 | 0.0 | 1.00 | 0.0 |
| 5 | $C_s$ | 1944.1 | -3.55 | 0.00933 | 0.00576 | 0.00433 | 0.0 | 0.98 | 0.20 |
| 5 | $C_{2v}$ | 1936.6 | -2.96 | 0.00824 | 0.00637 | 0.00628 | 1.00 | 0.0 | 0.0 |
| 6 | $C_s$ | 2448.1 | -4.34 | 0.00600 | 0.00516 | 0.00342 | 0.93 | 0.0 | 0.36 |
| 7 | $C_1$ | 2975.0 | -5.06 | 0.00451 | 0.00363 | 0.00283 | 0.09 | 0.84 | 0.53 |
| 8 | $C_{2v}$ | 3604.1 | -5.34 | 0.00349 | 0.00312 | 0.00251 | 0.0 | 0.0 | 1.00 |
| 9 | $C_{3v}$ | 4249.8 | -6.37 | 0.00277 | 0.00277 | 0.00211 | 0.0 | 0.0 | 1.00 |

[a] Sym = point group symmetry. Binding energies and rotational constants $A$, $B$, $C$, are in units of cm$^{-1}$. Dipole moment is normalized to unity.

Table 3. Calculated equilibrium structures of $CO_2$-$Xe_n$ clusters.[a]

| n | Sym | Binding energy | $A$ | $B$ | $C$ | $\mu_a$ | $\mu_b$ | $\mu_c$ |
|---|-----|---------|------|------|------|------|------|------|
| 3 | $C_s$ | 1241.3 | 0.01043 | 0.00982 | 0.00644 | 1.00 | 0.0 | 0.09 |
| 4 | $C_{2v}$ | 1846.9 | 0.00816 | 0.00466 | 0.00410 | 0.0 | 1.00 | 0.0 |
| 5 | $C_{2v}$ | 2488.8 | 0.00488 | 0.00364 | 0.00346 | 1.00 | 0.0 | 0.0 |
| 6 | $C_1$ | 3189.6 | 0.00369 | 0.00288 | 0.00221 | 0.48 | 0.29 | 0.83 |
| 7 | $C_1$ | 3881.2 | 0.00265 | 0.00243 | 0.00172 | 0.07 | 0.94 | 0.32 |
| 8 | $C_{2v}$ | 4681.3 | 0.00190 | 0.00177 | 0.00140 | 0.0 | 0.0 | 1.00 |
| 9 | $C_s$ | 5518.0 | 0.00179 | 0.00132 | 0.00115 | 0.27 | 0.0 | 0.96 |

[a] Sym = point group symmetry. Binding energies and rotational constants $A$, $B$, $C$, are in units of cm$^{-1}$. Dipole moment is normalized to unity.

Figure 1 illustrates the so-called chemical potential as a function of cluster size, given by the difference in binding energy between $CO_2$-$Rg_n$ and $CO_2$-$Rg_{n-1}$. This quantity thus represents the increase in binding energy due to the addition of Rg atom number n. The present results for $CO_2$-$Ar_n$ are essentially identical to those in Fig. 4 of Ref. 27. A larger magnitude for the chemical potential (i.e. a minimum in Fig. 1) indicates a particularly stable cluster, while a smaller magnitude indicates a less stable one. Thus, for example, $CO_2$-$Ar_{11}$ and -$Ar_{17}$ are calculated to be extra-stable compared to $CO_2$-$Ar_{12}$ and -$Ar_{16}$. We will see below that higher stability tends to be related to more symmetric cluster structures. The results in Fig. 1 show some interesting differences in the progression of the chemical potential between Ar, Kr, and Xe.

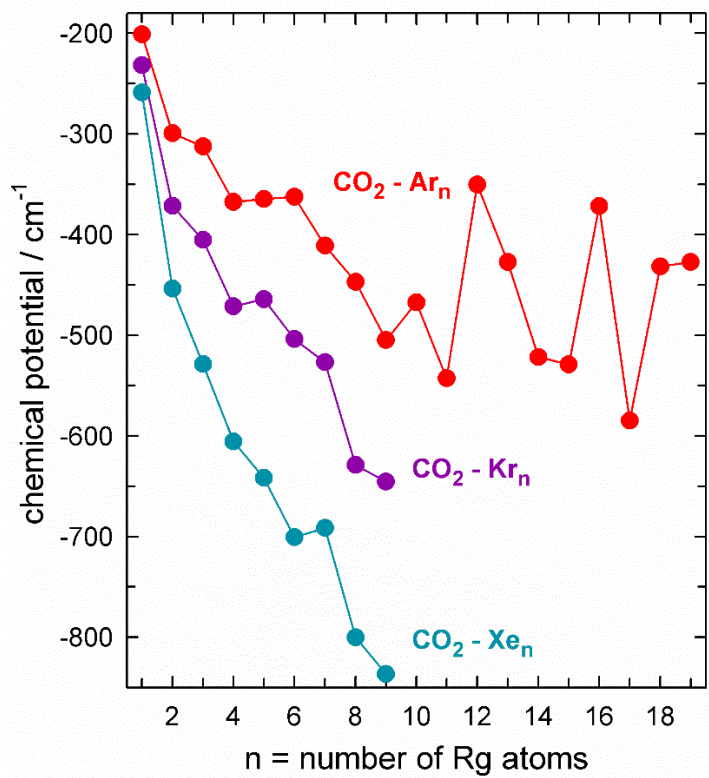

Figure 1. Calculated chemical potentials for $CO_2$-$Rg_n$ clusters, given by $BE(CO_2$-$Rg_{n-1}) - BE(CO_2$-$Rg_n)$, where BE = Binding Energy from Tables $1 - 3$. This is the binding energy gained by adding the nth Rg atom, and is thus a measure of cluster stability. The present results for $CO_2$-$Ar_n$ are essentially identical to those of Ref. 27.

## 4. Observed spectra

Spectra and structures for $CO_2$-$Rg_n$ clusters with n = 1 and 2 have been studied previously as described in the Introduction. As noted there, geometrical structures for $CO_2$-$Rg_2$ trimers are obvious in advance from those of the respective dimers, $CO_2$-Rg and Rg-Rg. But structures for larger clusters are not so obvious, and here we describe our theoretical and experimental results, starting with $CO_2$-$Rg_3$ and working up in size.

### 4.1. $CO_2$-$Rg_3$, Rg = Ar, Kr, Xe

One possible structure for $CO_2$-$Rg_3$ simply adds the third Rg atom in a position equivalent to the first two, continuing the formation of an equatorial ring. This first structure has $C_{2v}$ symmetry (assuming the Rg atoms are indistinguishable), with the symmetry axis passing through the C atom and the central Rg atom. A second possible structure takes an $Rg_3$ trimer, which is an equatorial triangle, and combines it with a $CO_2$ such that two of the Rg atoms are in near equatorial positions (like $CO_2$-$Rg_2$) and the third Rg atom is located towards one end of the $CO_2$, closer to an O atom. This gives $C_s$ symmetry, with a symmetry plane containing the $CO_2$ and the end Rg atom. These two $CO_2$-$Rg_3$ structures are illustrated in Fig. 2. The question of which structure has the lower energy depends on the relative strength of the $CO_2$-Rg and Rg-Rg interactions, as well as the shapes of the interaction potentials. In their original 1996 $CO_2$-$Ar_2$ paper, Sperhac et al.[20] favored the first ($C_{2v}$ 'ring') structure for $CO_2$-$Ar_3$. A calculation by Jose and Gadre[26] in 2007 agreed with $C_{2v}$, whereas Severson[24] in 1998, Boytuka et al.[25] in 2007, and Wang and Xie[27,28] in 2011-12 favored the second ($C_s$ 'end') structure for $CO_2$-$Ar_3$.

Our equilibrium global minimum calculations yielded the $C_{2v}$ structure for $CO_2$-$Ne_3$ and the $C_s$ structure for $CO_2$-$Ar_3$, -$Kr_3$, and -$Xe_3$. The actual energy differences ($C_{2v}$ minus $C_s$) were -13, +5, +25, and +61 cm$^{-1}$ for $CO_2$-$Ne_3$, -$Ar_3$, -$Kr_3$, and -$Xe_3$, respectively. Evidently the $CO_2$-Rg attraction dominates for $CO_2$-$Ne_3$, but the Rg-Rg attraction increasingly stabilizes the triangular $Rg_3$ structure, and thus the $C_s$ structure for $CO_2$-$Rg_3$ in the case of heavier Rg atoms. In this progression it is clear that $CO_2$-$Ar_3$ lies close to the point where the two structures crossover in energy.

The two structures predict very different spectral patterns, as illustrated for $CO_2$-$Ar_3$ in the upper left-hand panel of Fig. 3, which shows simulated spectra based on our calculated cluster structures. Here the rotational temperature is 2 K and the calculated equilibrium rotational constants are scaled by a factor of 0.97 to account for zero-point effects. Note the very prominent central $Q$-branch for the $C_s$ structure and the more spread out central intensity for the $C_{2v}$ structure. Our observed spectra, shown in the remaining panels of Fig. 3, unambiguously support the $C_s$ structures for $CO_2$-$Ar_3$, -$Kr_3$, and -$Xe_3$. As already noted, we have not been able to assign a $CO_2$-$Ne_3$ spectrum. Figure 3 illustrates that the observed absorption lines of $CO_2$-$Rg_3$ beyond the central $Q$-branch are relatively weak and partially obscured by lines of $CO_2$-Rg and -$Rg_2$.

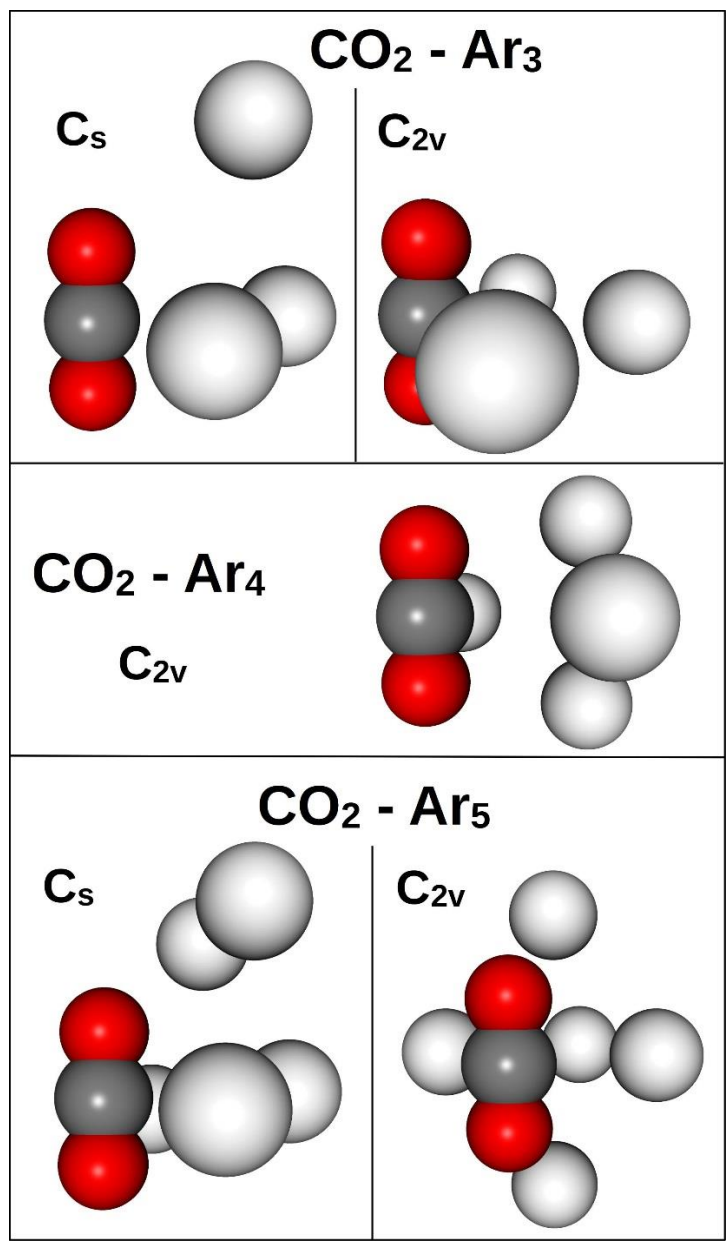

Figure 2. Calculated equilibrium structures of $CO_2$-$Ar_n$ clusters, n = 3 - 5. For $CO_2$-$Ne_3$, the $C_{2v}$ geometry is most stable, while for $CO_2$-$Ar_3$, -$Kr_3$, and -$Xe_3$, $C_s$ is most stable. For $CO_2$-$Ne_5$, -$Ar_5$, and -$Kr_5$, $C_s$ is most stable, while for $CO_2$-$Xe_5$, $C_{2v}$ is most stable.

For $CO_2$-$Xe_3$ there were 34 assigned lines, for $CO_2$-$Kr_3$ there were 25, and for $CO_2$-$Ar_3$ there were only 8. We fit these observations to obtain the parameters listed in Table 4. The ground and excited state rotational constants were constrained to be equal because of the relatively small number of assigned lines (especially for $CO_2$-$Ar_3$). All observed lines were blends of two or more individual transitions. Here and elsewhere we used the Mergeblends feature of Pgopher to fit each observation to an intensity weighted average of its underlying transitions. Due to the limited data, the parameters for $CO_2$-$Ar_3$ are not particularly well determined. Those for $CO_2$-$Kr_3$ are better, and those for $CO_2$-$Xe_3$ best. Table 4 includes for comparison the rotational constants predicted by our cluster structure calculations. As expected, the observed constants are slightly smaller than the predicted equilibrium ones due to zero-point effects.

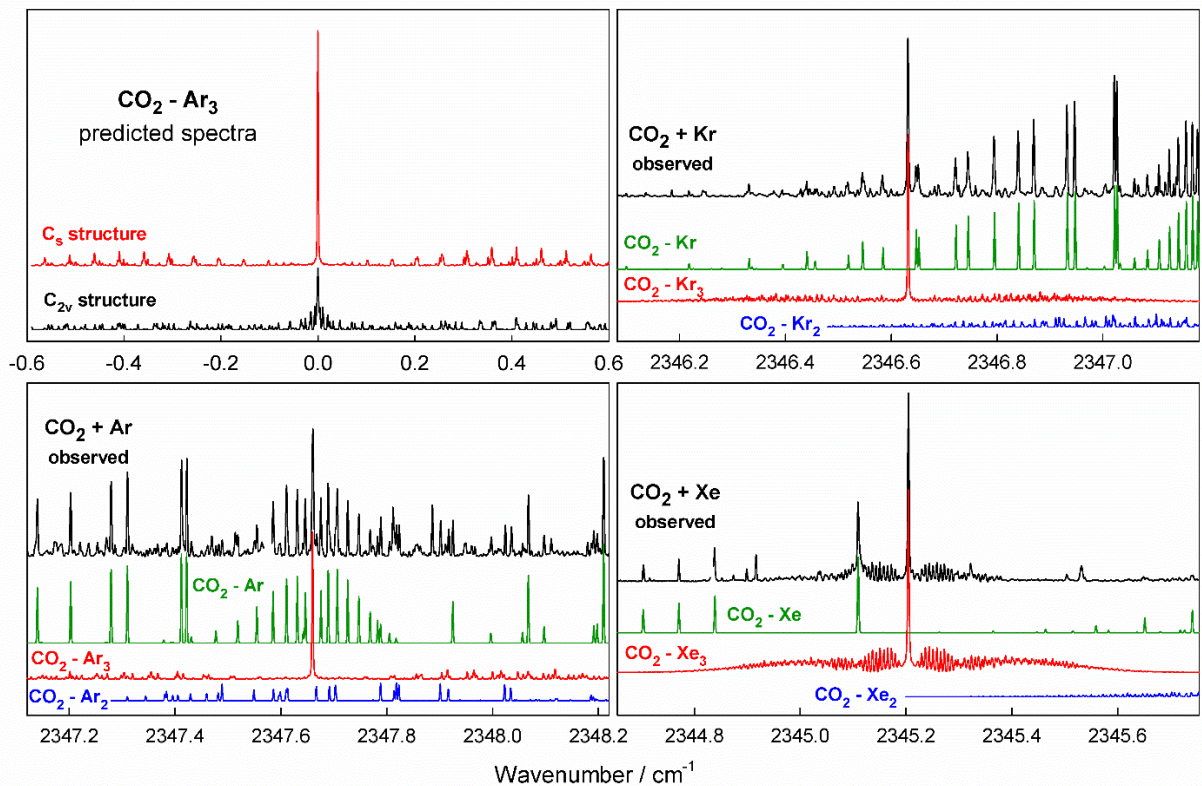

Figure 3. Simulated and observed spectra of $CO_2$-$Ar_3$, -$Kr_3$, and $Xe_3$. Gaps in the observed spectra are regions of $CO_2$ monomer absorption.

## 4.2. $CO_2$-$Rg_4$, Rg = Ar, Kr, Xe

The obvious minimum energy structure for an $Rg_4$ tetramer locates the Rg atoms at the vertices of a tetrahedron, since this allows six fully optimized Rg-Rg bonds. Our calculated minimum energy isomer for $CO_2$-$Rg_4$ (Rg = Ar, Kr, Xe) takes this tetrahedron structure and positions the $CO_2$ so that two of the Rg atoms are in equatorial positions (similar to $CO_2$-$Rg_2$), and the other two are pointing away, as shown in the lower part of Fig. 2. Of course the presence of $CO_2$ means that the Rg atoms no longer form a perfect tetrahedron. This structure has $C_{2v}$ symmetry, with the symmetry axis being the $b$-inertial axis for $CO_2$-$Ar_4$, or the $c$ axis for $CO_2$-$Kr_4$ and -$Xe_4$. The infrared transition moment (along the $CO_2$ axis) is $c$-type for $CO_2$-$Ar_4$, or $b$-type for $CO_2$-$Kr_4$ and -$Xe_4$.

Table 4. Molecular parameters for $CO_2$-$Rg_3$ (in $cm^{-1}$) [a]

| | $CO_2$-$Ar_3$ | | $CO_2$-$Kr_3$ | | $CO_2$-$Xe_3$ | |
|---|---|---|---|---|---|---|
| | Obs | Calc | Obs | Calc | Obs | Calc |
| $\nu_0$ | 2347.6598(8) | | 2346.6319(1) | | 2345.2044(1) | |
| $A$ | 0.0329(24) | 0.03448 | 0.01705(16) | 0.01728 | 0.010180(48) | 0.01043 |
| $B$ | 0.02660(21) | 0.02705 | 0.015124(63) | 0.01560 | 0.009816(30) | 0.00982 |
| $C$ | 0.02396(15) | 0.02413 | 0.011322(16) | 0.01154 | 0.006396(8) | 0.00644 |

[a] Quantities in parentheses correspond to $1\sigma$ from the least-squares fit, in units of the last quoted digit. Ground and excited state rotational parameters were constrained to be equal.

Our spectrum of $CO_2$-$Ar_4$ is shown in Fig. 4. Although the transitions are fairly weak and there are many interfering lines, the central region for $CO_2$-$Ar_4$ ($\approx$2347.05 – 2347.35 $cm^{-1}$) can be well simulated using parameters close to those expected from its calculated structure. The features in this region are $c$-type $Q$-branches, with a pair of distinctive central 'lumps' having $K_a = 0 \leftarrow 1$ and $1 \leftarrow 0$ (2347.173 and 2347.184 $cm^{-1}$). These are surrounded by $Q$-branches with $K_a = 1 \leftarrow 2$ and $2 \leftarrow 1$, $K_a = 2 \leftarrow 3$ and $3 \leftarrow 2$, and so on. We assigned 28 lines which were fit to obtain the parameters listed in Table 5. The simulation and fit assume $C_{2v}$ symmetry and thus include only ground state levels with $(K_a, K_c)$ = (even, even) and (odd, odd). But this makes almost no difference to the simulated spectrum and so does not in itself confirm the cluster symmetry.

As mentioned, the rotational selection rule predicted for $CO_2$-$Kr_4$ and -$Xe_4$ is $b$-type, and this results in a less distinctive band structure compared to $CO_2$-$Ar_4$. Our observed spectra are shown in Fig. 5, and it turns out that these $CO_2$-$Rg_4$ bands are overlapped by $CO_2$-$Rg_5$ (see below). The situation is clearer for Kr, shown in the upper panel of Fig. 5. The band center for $CO_2$-$Kr_4$ is located in the middle of the $R$-branch of $CO_2$-$Kr_5$, but there is no central $Q$-branch for $CO_2$-$Kr_4$ since this is a $b$-type band. The overlap confuses matters by blending together many of the $CO_2$-$Kr_4$ and -$Kr_5$ lines, but fortunately the $P$- and $Q$-branches of $CO_2$-$Kr_5$ are less obscured so its $R$-branch can be well predicted ($CO_2$-$Kr_5$ and -$Xe_5$ are described below).

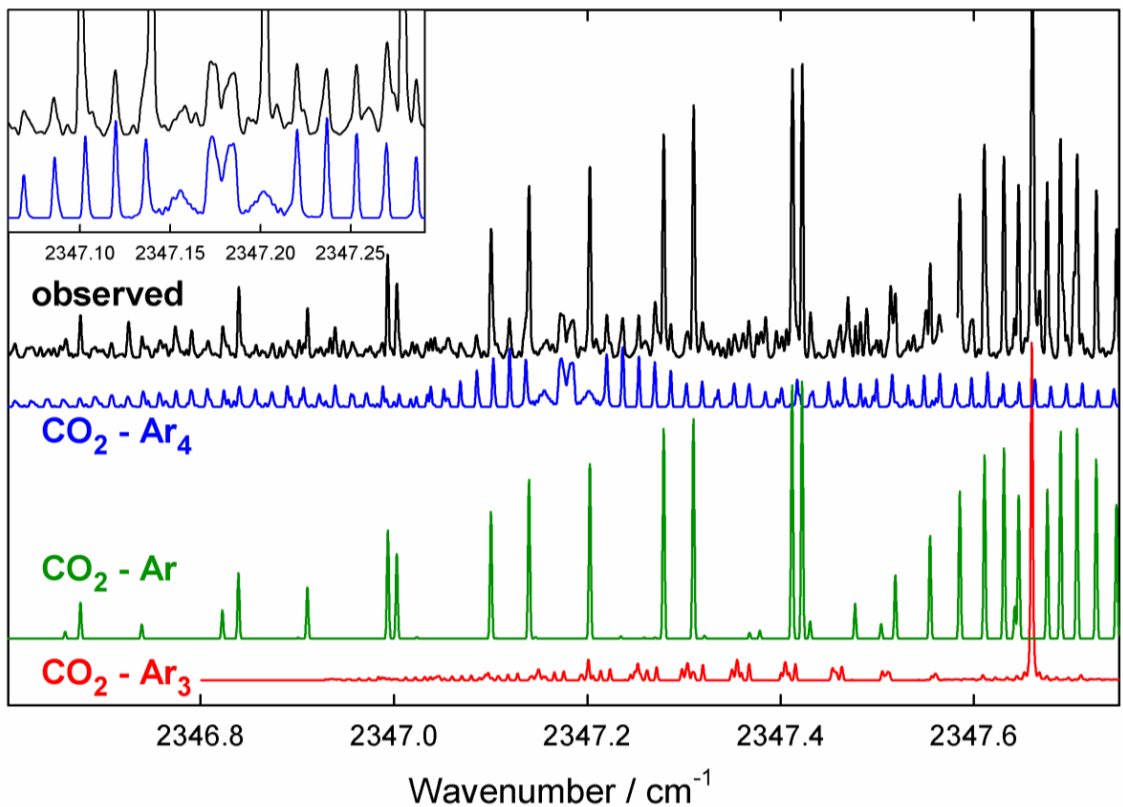

Figure 4. Simulated and observed spectra of CO₂-Ar₄. Gaps in the observed spectrum are regions of CO₂ monomer absorption. The inset shows the central $Q$-branch region of CO₂-Ar₄.

Table 5. Molecular parameters for CO₂-Rg₄ (in cm⁻¹) [a]

|  | CO₂-Ar₄ | | CO₂-Kr₄ | | CO₂-Xe₄ | |
|---|---|---|---|---|---|---|
|  | Obs | Calc | Obs | Calc | Obs | Calc |
| $\nu_0$ | 2347.1785(1) | | 2345.8765(1) | | 2344.066(30) | |
| $A'$ | 0.024739(11) | | | | | |
| $A''$ | 0.024763(10) | 0.02551 | 0.01378(13) | 0.01371 | [0.00817] | 0.00816 |
| $(B + C)/2$ | 0.016371(13) | 0.01712 | 0.00794(14) | 0.00779 | 0.00437 [b] | 0.00438 |
| $(B - C)$ | 0.00025 [b] | 0.00038 | [0.0006] | 0.00067 | [0.0005] | 0.00057 |

[a] Quantities in parentheses correspond to 1σ from the least-squares fit, in units of the last quoted digit. Ground and excited state rotational parameters were constrained to be equal except for $A$ of CO₂-Ar₄. Parameters in square brackets were fixed at scaled calculated values.

[b] Adjusted manually to fit the band shape.

The spectrum in the lower panel of Fig. 5 showing $CO_2$-$Xe_4$ and -$Xe_5$ is approximately scaled and aligned in order to show the analogy with $CO_2$-$Kr_4$ and -$Kr_5$ in the upper panel. However, the situation is less clear for $CO_2$-$Xe_4$ since its spectrum is weaker and less well-resolved. The Kr result suggested and supports the $CO_2$-$Xe_4$ assignment, which we think is persuasive but not absolutely conclusive. We analyzed 16 assigned lines of $CO_2$-$Kr_4$ in a highly constrained fit to obtain the parameters listed in Table 5 and used in the simulation in Fig. 5. A fit was not possible for $CO_2$-$Xe_4$, whose simulation is based on our calculated parameters, which were slightly adjusted manually and/or scaled to allow for zero-point effects. Note the rather large quoted uncertainty for the band origin of $CO_2$-$Xe_4$, which could possibly change up or down by one or two steps of about 0.0075 cm$^{-1}$.

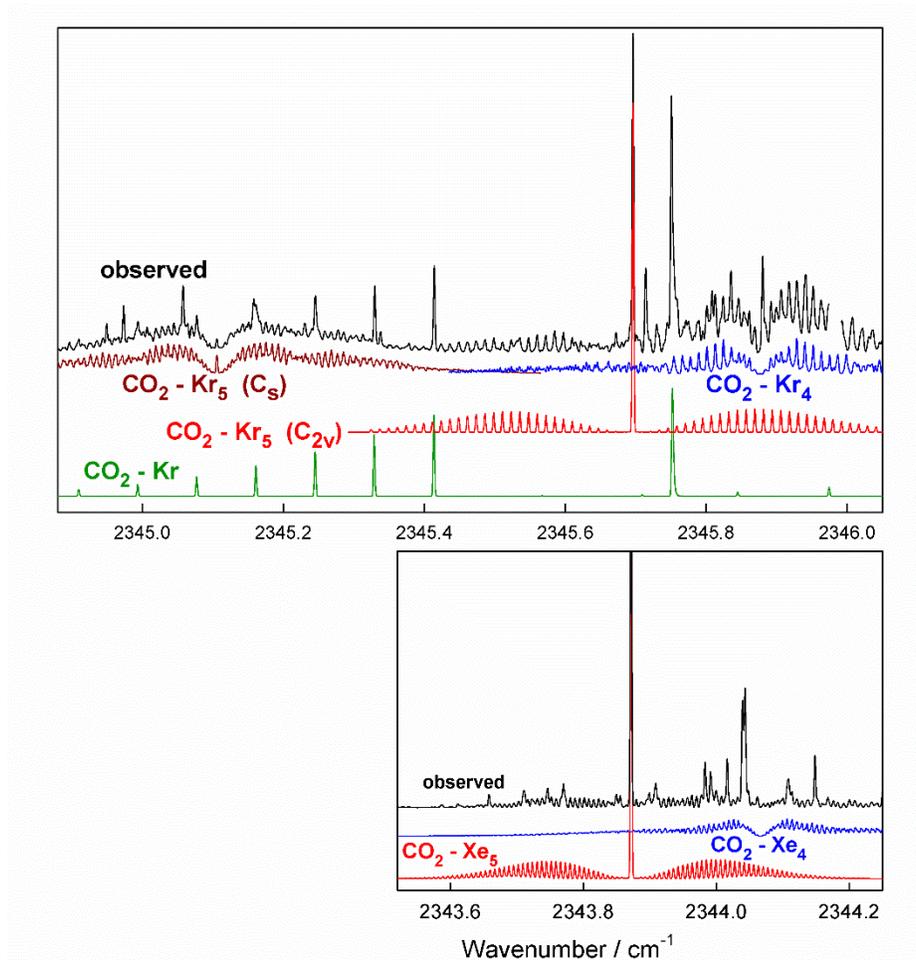

Figure 5. Simulated and observed spectra of $CO_2$-$Kr_4$, -$Kr_5$ (upper panel), -$Xe_4$, -$Xe_5$ (lower panel). The lower panel is aligned and scaled to illustrate the analogy between the Kr and Xe spectra. Gaps in the observed spectra are regions of $CO_2$ monomer absorption.

### 4.3. CO₂-Rg₅, Rg = Kr, Xe

With an increasing number of Rg atoms, the number of low-lying isomers increases rapidly. Our calculated CO₂-Rg₅ global minimum structure for Rg = Ne, Ar, and Kr has C$_s$ symmetry, while that for Xe has C$_{2v}$ symmetry. These two structures are illustrated in Fig. 2. The same C$_{2v}$ structure is the second most stable isomer for CO₂-Ar₅ and -Kr₅, while the same C$_s$ structure is the fifth most stable isomer for CO₂-Xe₅, with the energy differences (C$_{2v}$ minus C$_s$) being +40.0, +7.5, and -44.9 cm⁻¹ for Rg = Ar, Kr, and Xe, respectively. The C$_{2v}$ symmetry axis is *b* for Ar and Kr and *c* for Xe; the transition moment is *a*-type for all three clusters. The C$_s$ transition moment is mostly *b*-type with a smaller *c* component. For CO₂-Xe₅, the second most stable isomer has a different C$_s$ structure with an arrangement somewhat similar to the C$_{2v}$ structure having three Xe atoms in-plane and two out-of-plane. For CO₂-Ne₅, the second most stable isomer is a different C$_{2v}$ structure having all five Ne atoms in equatorial positions.

There are two relatively weak unexplained peaks (2346.514 or 2346.539 cm⁻¹) which are possible candidates for CO₂-Ar₅, but assignment is uncertain. Why are we able to assign CO₂-Kr₅ and -Xe₅ (see below), but not CO₂-Ar₅? This could be because the C$_{2v}$ isomer of CO₂-Ar₅ (with a more distinctive spectrum) is calculated to be 40 cm⁻¹ higher in energy than C$_s$ (whose predicted spectrum is not distinctive). Another factor is that vibrational shifts are smaller for Ar clusters, making CO₂-Ar₅ more difficult to spot because it is overlapped by other species.

We do have two distinct bands for CO₂-Kr₅ and one for CO₂-Xe₅, as already shown in Fig. 5. The predicted C$_{2v}$ isomer, a near-prolate slightly asymmetric rotor with *a*-type selection rules, provides an excellent match for the CO₂-Kr₅ and -Xe₅ bands at 2345.696 and 2343.872 cm⁻¹, respectively. We fit these bands assuming equal upper and lower state rotational constants and fixing *A* and (*B* - *C*) at values scaled from our calculated structures. The band origins and (*B* + *C*)/2 values were adjusted, with results as shown in Table 6. No spin statistical weights were used in the simulations since many if not most of these clusters will have different Kr or Xe isotopes in the equivalent positions, and in any case it makes almost no difference since *K*-structure is not resolved in our spectra.

Another band located at 2345.106 cm⁻¹ was assigned to the C$_s$ isomer of CO₂-Kr₅ (see Fig. 5). Again, it was well simulated using parameters very close to those expected from our calculated structure. The fit was highly constrained, similar to those in the last paragraph, and the results are shown in Table 6. This band is mostly *b*-type, with a small *a*-type component. The predicted relative transition moments are 0.98 and 0.20, respectively, and the weak $a$-type $Q$-branch is clearly visible in the observed and simulated spectra of Fig. 5, helping to confirm the assignment.

Table 6. Molecular parameters for $CO_2$-$Kr_5$ and $CO_2$-$Xe_5$ (in cm$^{-1}$) [a]

| | $CO_2$-$Kr_5$ ($C_s$) | | $CO_2$-$Kr_5$ ($C_{2v}$) | | $CO_2$-$Xe_5$ ($C_{2v}$) | |
|---|---|---|---|---|---|---|
| | Obs | Calc | Obs | Calc | Obs | Calc |
| $\nu_0$ | 2345.1062(1) | | 2345.6962(1) | | 2343.8720(1) | |
| $A$ | 0.009140(18) | 0.00933 | [0.00800] | 0.00824 | [0.00475] | 0.00488 |
| $(B + C)/2$ | 0.004917(9) | 0.00505 | 0.0061771(48) | 0.00633 | 0.0035338(33) | 0.00355 |
| $(B - C)$ | [0.00141] | 0.00143 | [0.00008] | 0.00009 | [0.00015] | 0.00018 |

[a] Quantities in parentheses correspond to 1σ from the least-squares fit, in units of the last quoted digit. Ground and excited state rotational parameters were constrained to be equal. Parameters in square brackets were fixed at scaled calculated values.

As mentioned above, our calculations for $CO_2$-$Kr_5$ put the $C_s$ isomer just 7.5 cm$^{-1}$ lower in energy than the $C_{2v}$ isomer, so it is not surprising that both structures are observed. From Fig. 5 one might expect to observe the analogous $C_s$ isomer of $CO_2$-$Xe_5$ to appear around 2343.2 cm$^{-1}$, but we do not have results for this range. Note, however, that this $C_s$ structure is 44.9 cm$^{-1}$ *higher* in energy than $C_{2v}$ for $CO_2$-$Xe_5$, suggesting that it is not so likely to be observed. Moreover, there are three other $CO_2$-$Xe_5$ isomers with energies calculated to lie between the $C_{2v}$ global minimum and the analogous $C_s$ structure.

In our recent studies of $CO_2$-Kr,[17] $CO_2$-Xe,[18] $CO_2$-$Kr_2$, and $CO_2$-$Xe_2$,[21] we explicitly included the effects of the various Kr and Xe isotopes. However for the n = 3, 4, and 5 clusters in the present work such a treatment becomes impractical and pointless. Thus the present simulations and fits simply assume that all atoms have the same average atomic mass of 84 or 131, respectively. One effect of this approximation is that lines with higher $J$-values in the simulated spectra remain sharp, while those in the observed spectra broaden and blur together because in reality the clusters have a range of different rotational constants arising from the different isotopes. As a result, for example, the positions of peak intensity in the $P$-branches of $CO_2$-$Kr_5$ and -$Xe_5$ (Fig. 5) occur at higher $J$-values in the simulations than in the observed spectra.

### 4.4. CO₂-Ar₆, -Ar₇, and -Ar₈

Our calculated equilibrium structure for $CO_2$-$Ar_6$ is shown in Fig. 6. It has $C_s$ symmetry, and resembles $CO_2$-$Ar_4$ with two additional Ar atoms added at one end (the top of the picture in Fig. 6). This structure predicts a mostly *a*-type band with a small *c*-type component. Similar structures are predicted for $CO_2$-$Kr_6$ and -$Xe_6$, but so far we have not assigned spectra of any clusters with n > 5 for Kr or Xe. We tentatively assign $CO_2$-$Ar_6$ to a band featuring a strong central *Q*-branch peak at 2346.1306 cm⁻¹ (not shown). Using scaled rotational constants, the simulated spectrum in the *P*- and *R*-branch regions is consistent with the observed spectrum, but it is not possible to perform a meaningful fit.

Calculated global minimum structures for $CO_2$-$Ar_7$ and -$Ar_8$ are also shown in Fig. 6. The former has an unsymmetrical geometry with three approximately equatorial Ar atoms and four more towards one end. The latter, $CO_2$-$Ar_8$, has a symmetry plane ($C_s$ point group) and a structure somewhat similar to $CO_2$-$Ar_7$ with three equatorial Ar atoms plus five more at one end. Both are fairly asymmetric rotors and have predominantly *b*-type selection rules plus weaker *c*-type components (and even weaker *a*-type for $CO_2$-$Ar_7$). As a result, their predicted spectra have modest central *Q*-branch peaks and are otherwise rather blurred and featureless. We do not have definite assignments for either of these clusters, but on the basis of vibrational shifts (see Sec. 5, below) it is plausible to associate them with unexplained peaks in the observed spectrum at 2345.735 cm⁻¹ for $CO_2$-$Ar_7$ and 2345.597 cm⁻¹ for $CO_2$-$Ar_8$. But note that there are additional weaker unexplained peaks at 2344.483, 2344.385, and 2345.768 cm⁻¹.

### 4.5. CO₂-Ar₉

$CO_2$-$Ar_9$ is calculated to have a symmetrical structure shown in Fig. 6, with three rings each containing three equivalent Ar atoms, giving $C_{3v}$ symmetry. The threefold rotational symmetry axis coincides with the O-C-O axis. This structure can be described as a 3-3-3 cap. This symmetric top cluster provides an excellent match to a band centered at 2345.309 cm⁻¹ shown in Fig. 7. The well determined experimental *B*-value of 0.00607 cm⁻¹ compares with a value of 0.00629 cm⁻¹ from the calculated structure. The ratio of 0.965 is just what we expect for zero-point and equilibrium values, as mentioned in Sec. 3. Note that for symmetric rotor molecules like $CO_2$-$Ar_9$ (and $CO_2$-$Ar_{11}$, -$Ar_{15}$, -$Ar_{17}$, below) the spectra are not sensitive to the value of the constant (*A* or *C*) for rotation around the symmetry axis.

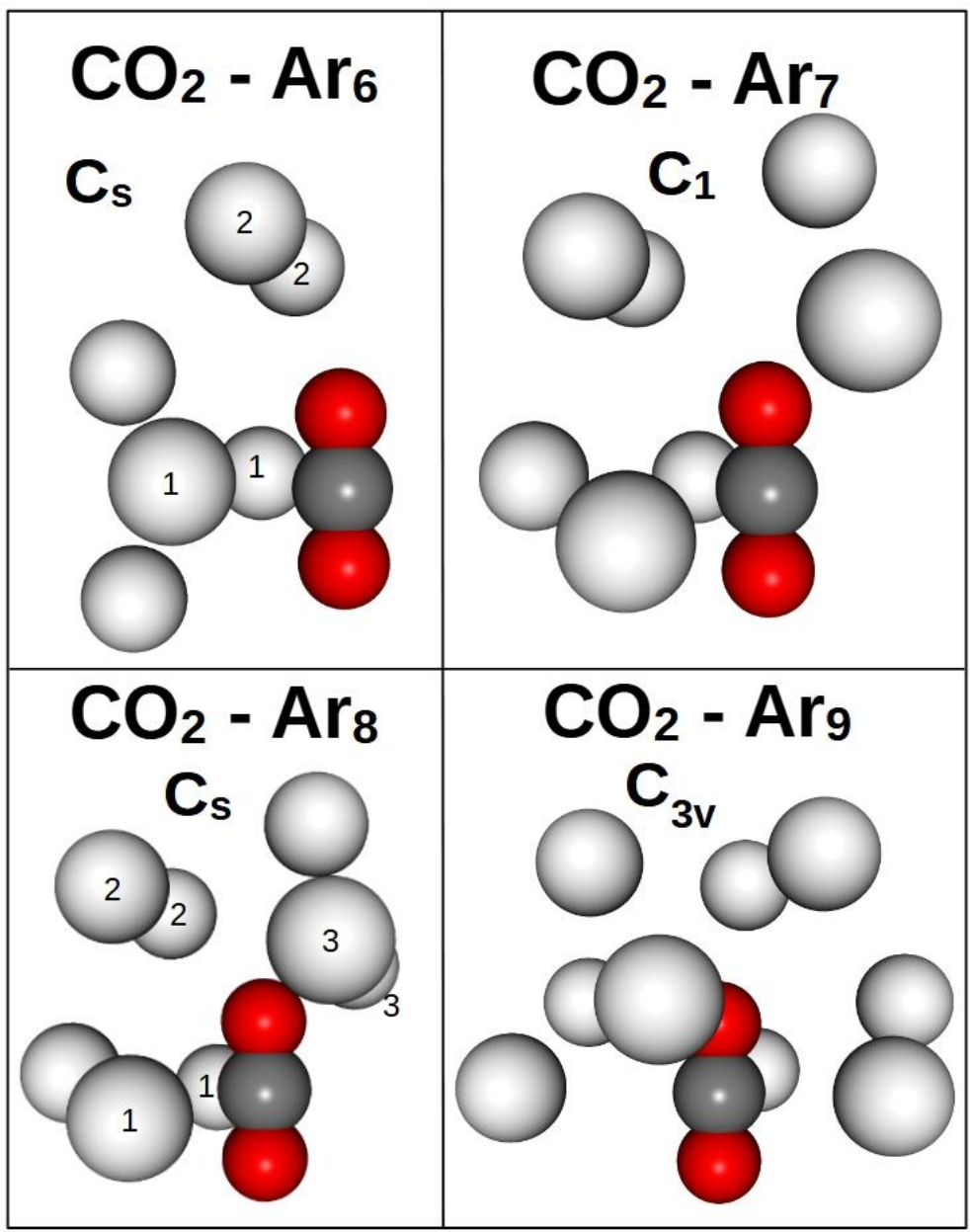

Figure 6. Calculated equilibrium structures of $CO_2$-$Ar_n$ clusters, n = 6 - 9. The symmetry plane for $CO_2$-$Ar_6$ contains the $CO_2$ molecule and the two leftmost Ar atoms. That for $CO_2$-$Ar_8$ contains the $CO_2$, the leftmost Ar, and the topmost Ar. In $CO_2$-$Ar_9$, the $CO_2$ molecule has a cap composed of 3 rings each containing 3 Ar atoms.

In addition to the main $CO_2$-$Ar_9$ $Q$-branch peak at 2345.309 cm$^{-1}$, there are two weaker peaks at 2345.318 and 2345.326 cm$^{-1}$ (and perhaps more beyond this) with relative intensities of about 0.43 and 0.19 (see inset to Fig. 7). There are also series of weaker peaks in the $P$- and $R$-branch regions, especially around 2344.95 to 2345.10, and 2345.35 to 2345.55 cm$^{-1}$. These weaker peaks can be well explained by associating them with the 2345.318 cm$^{-1}$ $Q$-branch feature and by using a slightly smaller $B$-value of 0.00591 cm$^{-1}$. Continuing the trend, we further assumed that the 2345.326 cm$^{-1}$ $Q$-branch peak belonged to a band with $B \approx 0.00575$ cm$^{-1}$. The three simulated bands and their sum are compared with the experimental spectrum in Fig. 7, and their parameters are listed in Table 7. The simulated sum provides an excellent match to experiment. For example, the fact that the $P$-branch lines of all three bands happen to align around 2345.18 cm$^{-1}$ explains why these lines appear stronger than any in the $R$-branch. We believe that the three apparent bands for CO$_2$-Ar$_9$ can be explained as a sequence arising from a very low frequency intermolecular vibrational mode, an idea which is explored further in Sec. 6 below. It should be emphasized that the assignment of the present band to CO$_2$-Ar$_9$ remains convincing even if this sequence band idea is not entirely correct.

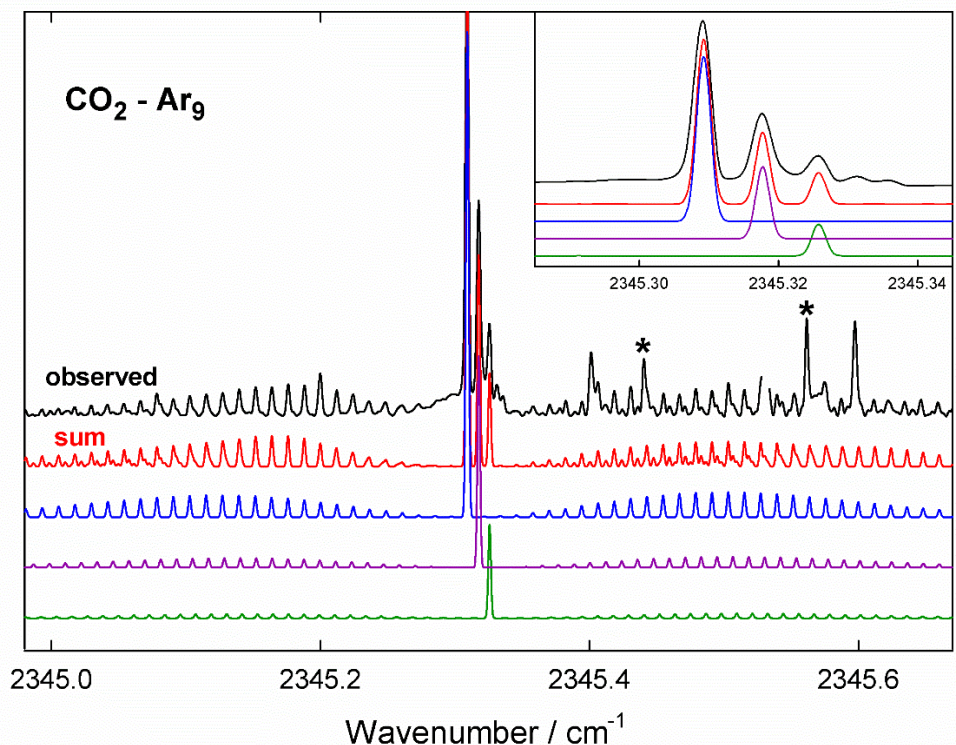

Figure 7. Simulated and observed spectra of CO$_2$-Ar$_9$. A gap in the observed spectrum is a region of CO$_2$ monomer absorption. Asterisks mark known CO$_2$-Ar dimer lines. There are three (or more) apparent $Q$-branch peaks (shown in the expanded inset) which may be a sequence arising from a low frequency ($\approx$2 cm$^{-1}$) thermally populated intermolecular vibrational mode (see text).

### 4.6. CO$_2$-Ar$_{10}$

The calculated global minimum structure of CO$_2$-Ar$_{10}$, shown in Fig. 8, has C$_{2v}$ symmetry. The twofold rotational axis coincides with the O-C-O axis and the $c$ inertial axis. The Ar atoms appear to have a 2-2-4-2 arrangement of rings, counting from the top in Fig. 8, but in fact the ring of 4 is actually composed of two rings of 2 with slightly different latitudes, so the arrangement is really 2-2-2-2-2. We assign $CO_2$-$Ar_{10}$ to a band centered at 2344.632 cm$^{-1}$ which is illustrated in the upper panel of Fig. 9. This assignment is quite consistent with the simulation shown, which is a *c*-type band with the following parameters, as scaled from the calculated structure: $A$, $B$, $C$ = 0.00557, 0.00464, 0.00418 cm$^{-1}$. However, the contour of the band is not well resolved in the *P*- and *R*-branch regions, and it was not possible to perform a meaningful fit to determine experimental rotational parameters.

Table 7. Molecular parameters for $CO_2$-$Ar_9$ (in cm$^{-1}$) [a]

|  | Main band | | Second band | Third band |
|---|---|---|---|---|
|  | Obs | Calc | Obs | Obs |
| $\nu_0$ | 2345.3095(1) |  | 2345.3178(1) | 2345.325 |
| $C$ | [0.00482] | 0.00492 | [0.00482] | [0.00482] |
| $B'$ | 0.0060669(7) |  | 0.0059091(13) | 0.00575 |
| $B''$ | 0.0060685(6) | 0.00629 | 0.0059096(12) | 0.00575 |

[a] Quantities in parentheses correspond to 1σ from the least-squares fit, in units of the last quoted digit. *C* values were fixed at a scaled calculated value. *B* for the third band was extrapolated from those of the first two bands.

### 4.7. $CO_2$-$Ar_{11}$

$CO_2$-$Ar_{11}$ has a calculated structure (Fig. 8) with $C_{5v}$ symmetry having two rings of 5 Ar atoms plus one Ar located at the top end on the O-C-O axis. We call this a 1-5-5 cap structure. The chemical potential calculations (Fig. 1) show that $CO_2$-$Ar_{11}$ and -$Ar_9$ are particularly stable, a fact which can be associated with their symmetric geometries. The $CO_2$-$Ar_{11}$ structure explains perfectly the band centered at 2344.0305 cm$^{-1}$ shown in the central panel of Fig. 9, and a fit to the observed spectrum gives the parameters listed in Table 8. Note that the scaling factor between observed and calculated *B*-values is 0.96, very similar to those of $CO_2$-$Ar_9$ and $Ar_2$. Our simulated spectra for $CO_2$-$Ar_{11}$ and -$Ar_9$ (above) include the effects of nuclear spin statistics, but in practice this makes almost no difference since *K*-structure was not resolved. $CO_2$-$Ar_{11}$ represents one of our most secure $CO_2$-$Ar_n$ assignments, along with those for n = 3, 4, 9, 15, and 17.

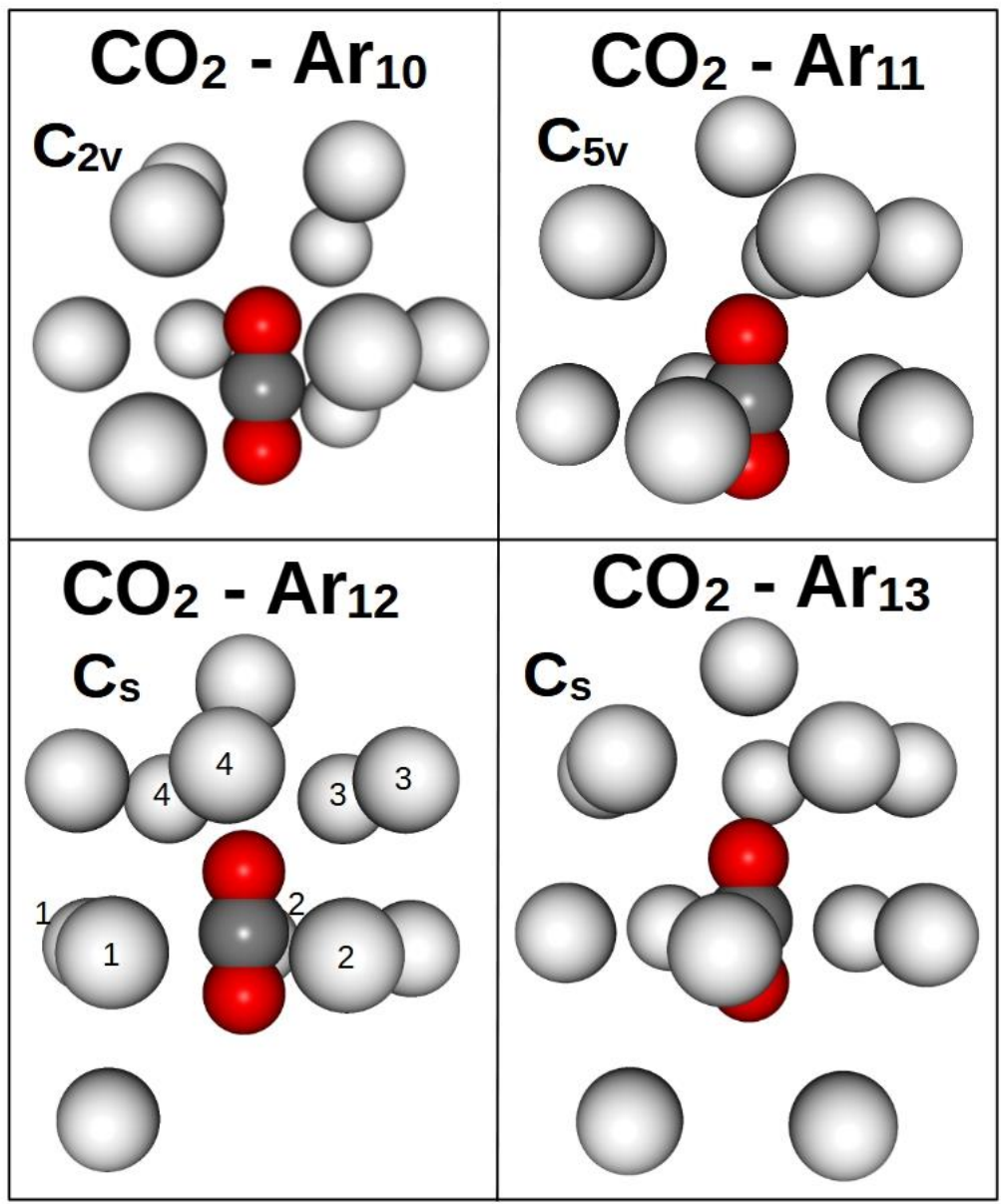

Figure 8. Calculated equilibrium structures of $CO_2$-$Ar_n$ clusters, n = 10 - 13. In $CO_2$-$Ar_{11}$, the $CO_2$ molecule has a cap composed a single Ar atom at the top, plus 2 rings each containing 5 Ar. The symmetry plane for $CO_2$-$Ar_{12}$ contains the $CO_2$ molecule and the four unlabeled Ar atoms; the labels indicate equivalent Ar atoms. The symmetry plane for $CO_2$-$Ar_{13}$ contains the $CO_2$ and 5 Ar atoms, including the topmost and the bottom two.

### 4.8. $CO_2$-$Ar_{12}$ and -$Ar_{13}$

The lowest energy calculated structures for $CO_2$-$Ar_{12}$ and -$Ar_{13}$ begin with highly symmetric $CO_2$-$Ar_{11}$ and then add one or two Ar atoms at the "bottom" (uncapped) end, as shown in Fig. 8. Each of these is a prolate asymmetric rotor with $C_s$ symmetry, and each has predominantly $a$-type transitions with a weaker $c$-type component. $CO_2$-$Ar_{12}$ is assigned to a band with a strong *Q*-branch at 2343.519 cm$^{-1}$ as shown in Fig. 9, and the fitted parameters are given in Table 9.

Assignment of $CO_2$-$Ar_{13}$ is more problematical. It could be responsible for a series of *P*- or *R*-branch peaks observed in the region near 2343.13 to 2343.20 cm$^{-1}$ (Fig. 9), but it is not clear where the band center lies. Possible *Q*-branch peaks could be those at 2343.009, 2343.057, 2343.114, or 2343.339 cm$^{-1}$. The calculated structures of $CO_2$-$Ar_{11}$, -$Ar_{12}$, and -$Ar_{13}$ suggest that the vibrational shift between n = 13 and 12 should be similar to that between 12 and 11, which would seem to favor the assignment of $CO_2$-$Ar_{13}$ to 2343.009 cm$^{-1}$. But simulations were inconclusive and for the moment we leave $CO_2$-$Ar_{13}$ as unassigned.

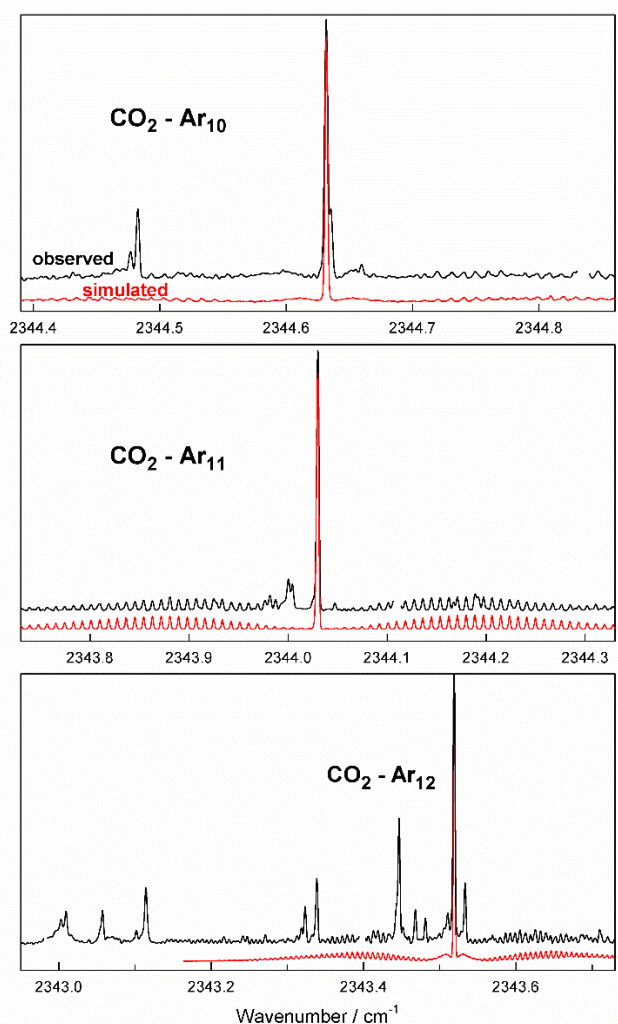

Figure 9. Simulated and observed spectra of $CO_2$-$Ar_{10}$, -$Ar_{11}$, and -$Ar_{12}$. Gaps in the observed spectra are regions of $CO_2$ monomer absorption. Weak fine structure just below 2343.2 cm$^{-1}$ may be due to $CO_2$-$Ar_{13}$ but a satisfactory assignment has not been established.

Table 8. Molecular parameters for $CO_2$-$Ar_{11}$, -$Ar_{15}$, and -$Ar_{17}$ (in $cm^{-1}$) [a]

|  | $CO_2$-$Ar_{11}$ | | $CO_2$-$Ar_{15}$ | | $CO_2$-$Ar_{17}$ | |
|---|---|---|---|---|---|---|
|  | Obs | Calc | Obs | Calc | Obs | Calc |
| $\nu_0$ | 2344.0305(1) | | 2341.9804(1) | | 2340.4719(1) | |
| $A$ or $C$ | [0.00386] | 0.00305 | [0.00314] | 0.00314 | [0.00260] | 0.00270 |
| $B' - B''$ | -0.00000140(7) | | -0.00000109(4) | | 0.0 | |
| $B''$ | 0.0043988(4) | 0.00458 | 0.0023325(8) | 0.00240 | 0.0018185(3) | 0.00189 |

[a] Quantities in parentheses correspond to 1σ from the least-squares fit, in units of the last quoted digit. Parameters in square brackets were fixed at scaled calculated values. The $CO_2$-$Ar_{15}$ and -$Ar_{17}$ results are taken from Ref. 4.

Table 9. Molecular parameters for $CO_2$-$Ar_{12}$ (in $cm^{-1}$) [a]

|  | Obs | Calc |
|---|---|---|
| $\nu_0$ | 2343.5191(1) | |
| $A$ | [0.00400] | 0.00412 |
| $(B + C)/2$ | 0.0033400(20) | 0.00344 |
| $(B - C)$ | [0.00014] | 0.00014 |

[a] Quantities in parentheses correspond to 1σ from the least-squares fit, in units of the last quoted digit. Parameters in square brackets were fixed at scaled calculated values.

### 4.9. $CO_2$-$Ar_{14}$ to -$Ar_{17}$

Our calculated minimum energy structures for $CO_2$-$Ar_{14}$ to -$Ar_{17}$ are shown in Fig. 10. These clusters have cage structures which essentially represent completion of the first solvation shell for argon around carbon dioxide. Both $CO_2$-$Ar_{14}$ and $CO_2$-$Ar_{16}$ have $C_s$ symmetry, and their symmetry planes lie almost in the plane of Fig. 10, with just a slight rotation so that all Ar atoms are visible. $CO_2$-$Ar_{15}$ and $CO_2$-$Ar_{17}$ have highly symmetric $D_{3h}$ and $D_{5h}$ structures, with the Ar atoms located respectively in 5 rings of 3 (3-3-3-3-3), and 3 rings of 5 plus 1 at each end (1-5-5-5-1).

The band assigned to $CO_2$-$Ar_{15}$ at 2341.980 $cm^{-1}$ and that assigned to $CO_2$-$Ar_{17}$ at 2340.472 $cm^{-1}$ are shown in Fig. 11. Their detailed observation and analysis were reported in our previous paper,[4] and the resulting parameters are summarized here in Table 8. The symmetry of these clusters and resulting nuclear spin statistics result in clear intensity alternation in both observed spectra (see Figs. 3 and 4 of Ref. 4), helping to make these assignments especially secure despite the relatively large cluster sizes.

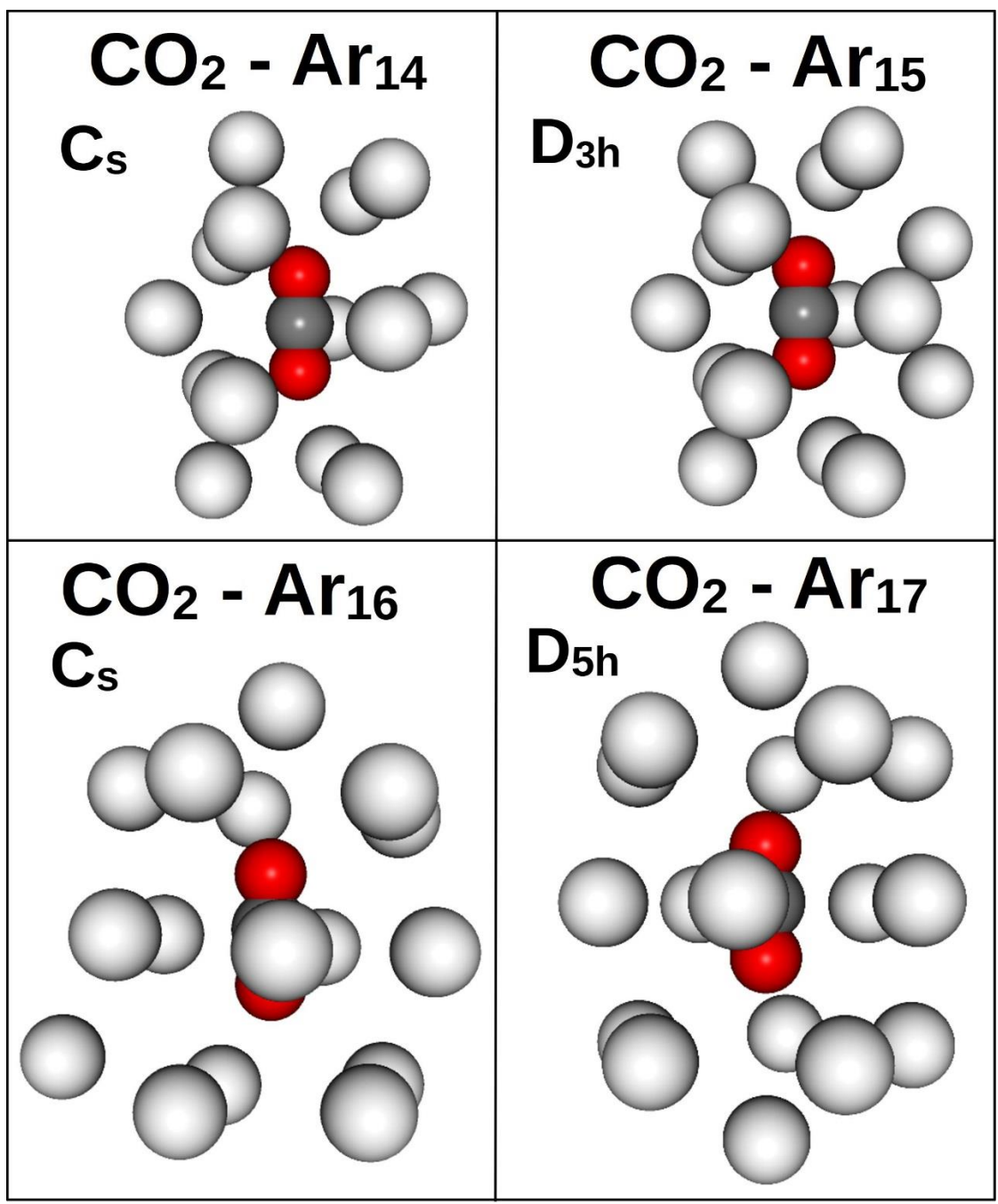

Figure 10. Calculated equilibrium structures of $CO_2$-$Ar_n$ clusters, n = 14 - 17. The symmetry planes for $CO_2$-$Ar_{14}$ and -$Ar_{16}$ are almost coincident with the plane of the drawing, with slight rotations so that all atoms are visible. $CO_2$-$Ar_{14}$ and -$Ar_{16}$ have symmetric cage structures with rings containing 3-3-3-3-3 and 1-5-5-5-1 Ar atoms, respectively.

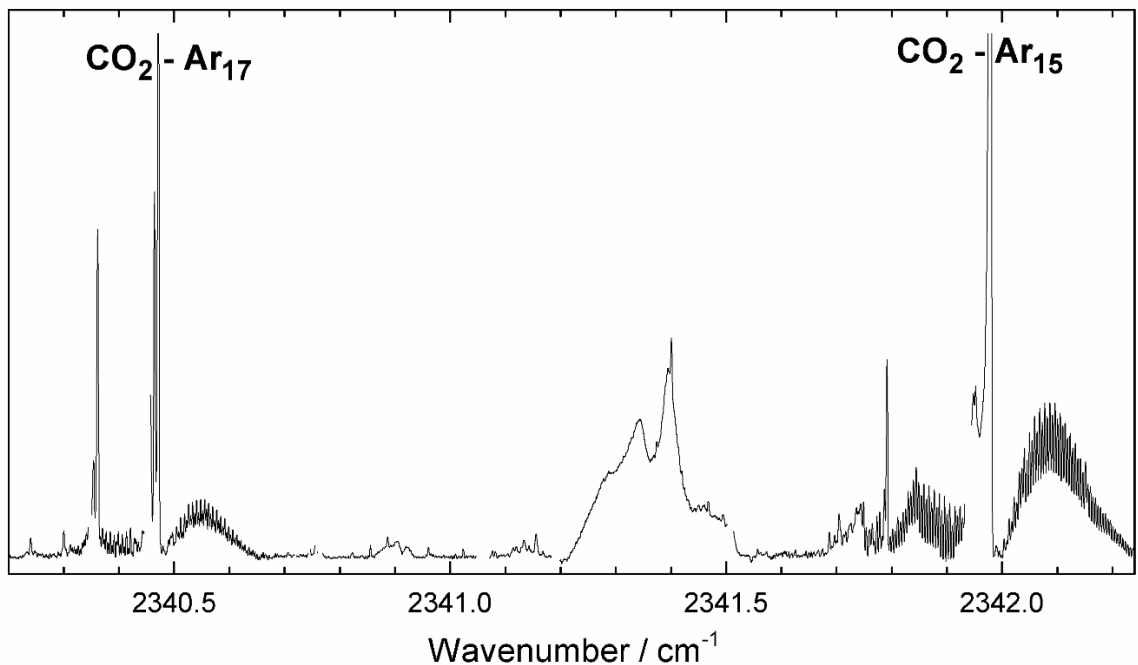

Figure 11. Observed spectrum showing bands assigned to $CO_2$-$Ar_{15}$ and -$Ar_{17}$ (see Ref. 4 for details) and other unassigned features due to $CO_2$-$Ar_n$ clusters. Gaps in the observed spectrum are regions of $CO_2$ monomer absorption.

We do not have an assignment for $CO_2$-$Ar_{14}$, but possible candidates are the unassigned $Q$-branches mentioned in the previous section, visible in the bottom panel of Fig. 9. The predicted vibrational shift for $CO_2$-$Ar_{14}$ is relatively small in magnitude (see Sec. 5 below), suggesting that it could even lie as high as the unidentified peaks at 2344.477 and 2344.483 cm$^{-1}$ visible in the top panel of Fig. 9.

$CO_2$-$Ar_{16}$ represents a peak in the $CO_2$-$Ar_n$ chemical potential curve (see Fig. 1), meaning that it is relatively less strongly bound than neighboring clusters. Another such peak occurs for n = 12, while n = 9, 11, 14, 15, and 17 represent relatively stable minima in the curve. Furthermore, our cluster calculations had some difficulty in converging on the exact global minimum for $CO_2$-$Ar_{16}$, and located 4 other local minima within 30 cm$^{-1}$. For comparison, most clusters with 8 < n < 19 had no other local minima within 30 cm$^{-1}$ (there were 2 for n = 12, and 5 for n = 13). These observations suggest that $CO_2$-$Ar_{16}$ has a particularly floppy or ill-defined structure which in turn could make its spectrum difficult to recognize. One possible assignment for $CO_2$-$Ar_{16}$ is the peak at 2341.792 cm$^{-1}$ (see Fig. 11). Note also the interesting region of strong but broad and unresolved absorption around 2341.2 to 2341.5 cm$^{-1}$. Additional relatively sharp unassigned features visible in Fig. 11 include: 2340.354, 2340.362, (these are possible candidates for $CO_2$-$Ar_{18}$), 2340.887, 2340.961, 2341.401, and 2341.792 cm$^{-1}$.

## 5. Vibrational shifts

The vibrational shifts of the cluster band origins relative to the free $CO_2$ frequency are a key aspect of the present results, since they are unambiguous and precisely determined (assuming the assignments are correct), as well as being amenable to theoretical calculation. Moreover, the shifts are of obvious interest in exploring non-additive intermolecular force effects, as well as for the interpretation of matrix isolation spectra of $CO_2$ in solid argon, where two lines are observed with characteristic shifts of -4 and -10 cm$^{-1}$.[39] The present shift results are shown graphically in Fig. 12. For n = 1 to 4, the shifts are quite linear for $CO_2$-Ar$_n$, -Kr$_n$, and -Xe$_n$ (left-hand panel of Fig. 12), though of course the small deviations from linearity are themselves of interest. For n = 5, the $C_s$ isomer of Kr continues the linear trend while the $C_{2v}$ isomers of Kr and Xe deviate significantly. Our calculated shifts for $CO_2$-Kr$_n$ consistently underestimate the magnitude of the observed shifts, but reproduce the pattern quite well. Note especially that the different shifts of the $C_s$ and $C_{2v}$ isomers of $CO_2$-Kr$_5$ are well predicted, helping to support the assignments.

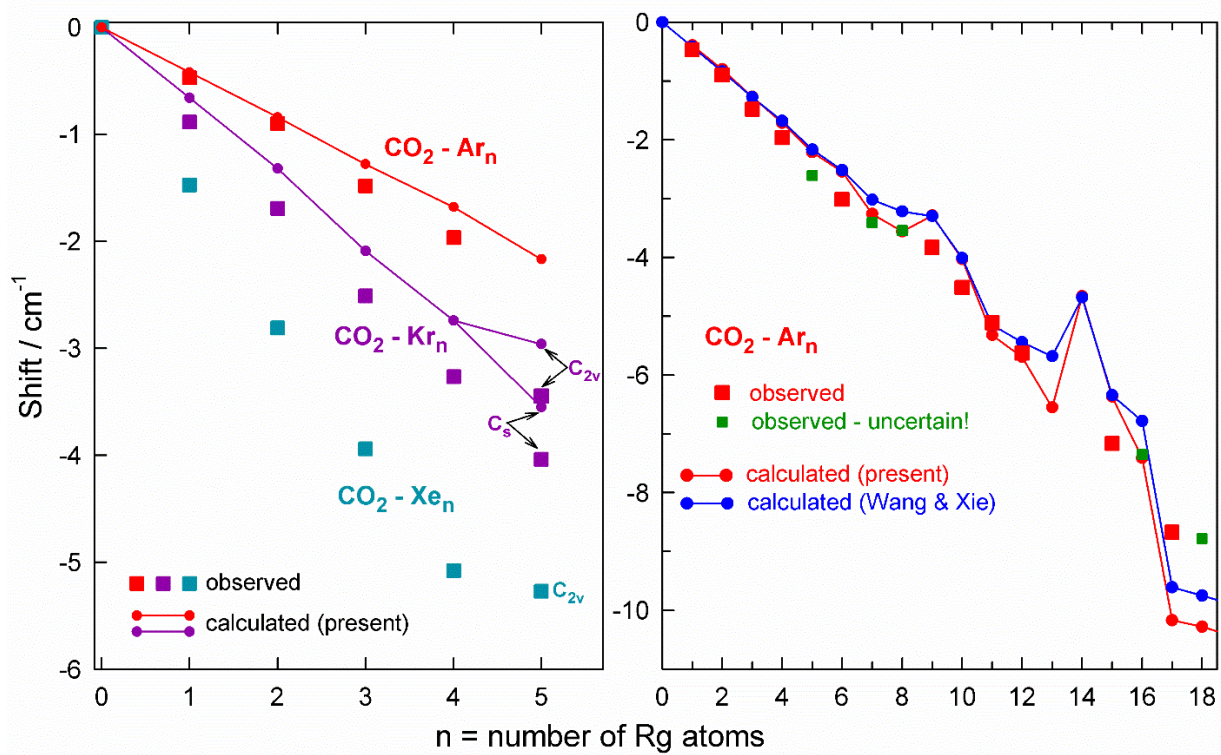

Figure 12. Observed and calculated vibrational frequency shifts for $CO_2$-Rg$_n$ clusters.

Beyond n = 5, we only have experimental results for $CO_2$-Ar$_n$ as shown in the right-hand panel of Fig. 12. The present calculated shifts in the figure represent the difference between our calculated equilibrium binding energies for the given cluster in the excited ($\nu_3$) and ground $CO_2$ vibrational states. As such, they rely on the *ab initio* potential surfaces of Cui et al.,[31] as fitted by us. The calculated shifts of Wang and Xie[28] also use the Cui et al. potentials, but they were performed using a path integral Monte Carlo method which should include the true zero point vibrational motions, at least to some extent. Severson[24] also calculated $CO_2$-$Ar_n$ vibrational shifts using a quantum Monte Carlo method and the results (not shown here) are fairly similar up to n = 15. However, those results assumed an excited ($\nu_3$) state surface that was simply scaled from that of the ground state, because a real exited state potential was not then available, and so we consider them to be less relevant.

The equilibrium and Monte Carlo calculated shifts in Fig. 12 follow roughly the same trend and are not so different from each other. The experimental results also follow the same general trend of negative shifts with a slope similar to that established for n = 1 − 4. The calculated shifts mostly tend to slightly underestimate the magnitude of the observed shifts. But interestingly the progression of the experimental points is smoother (more nearly linear) and seems not to follow fully the 'kinks' in the calculated points. These kinks are related to structural changes in the clusters, as nicely explained by Wang and Xie.[28] Note in particular that the calculations underestimate the magnitude of the shift for $CO_2$-$Ar_{15}$ and overestimate it for $CO_2$-$Ar_{17}$. As we said previously,[4] this may be due to the significant rearrangement of Ar atoms between n = 15 to 17 (Fig. 10) combined with shortcomings in the calculated potential surfaces of Cui et al.[31]

## 6. Sequence bands

In Fig. 7 and Section 4.5, we noted the appearance of a sequence of at least three *Q*-branch peaks belonging to $CO_2$-$Ar_9$. A similar sequence is present in the spectrum of $CO_2$-$Ar_{17}$ as reported in Ref. 4, and the progression extends to lower frequencies, rather than higher as for $CO_2$-$Ar_9$. In addition, for $CO_2$-$Ar_{15}$ there is at least one barely resolved low frequency *Q*-branch shoulder, and for $CO_2$-$Ar_{10}$ one high frequency *Q*-branch shoulder, which could belong to similar sequences. Furthermore, the unassigned peaks at 2344.483 and 2340.462 cm⁻¹ have possible sequence companions. We believe that these sequences can be explained as due to very low-frequency ($\approx$2 cm⁻¹) intermolecular vibrational modes of the clusters, as already mentioned briefly in Ref. 4.

Observation of a sequence depends on resolving its peaks separately from each other, which occurs because they have different vibrational shifts. In the case of $CO_2$-$Ar_9$, each successive peak moves up by about 0.008 cm$^{-1}$ (Fig. 7), meaning that each excitation of the intermolecular mode evidently reduces the magnitude of the vibrational shift by about 0.2 %. In the case of $CO_2$-$Ar_{17}$, each successive peak moves down by about 0.007 cm$^{-1}$ (Fig. 4 of Ref. 4), meaning that excitation increases the magnitude of the shift by about 0.08 %. For $CO_2$-$Ar_{15}$ the observed $Q$-branch shoulder is shifted down by approximately 0.004 cm$^{-1}$. For $CO_2$-$Ar_{11}$ and -$Ar_{12}$ where no $Q$-branch splitting is observed, it is still possible that there could be sequences present in the spectrum which remain unresolved because there are only small changes in vibrational shift between their components.

The decrease in the intensity of successive sequence $Q$-branch peaks is a straightforward thermal (Boltzmann) population effect, and in principle these relative intensities allow determination of the energy of the supposed low-frequency vibrational mode. In Ref. 4 we estimated this energy for $CO_2$-$Ar_{17}$ to be 1.3 cm$^{-1}$ if the mode is singly degenerate, or 2.4 cm$^{-1}$ if doubly degenerate, but there is considerable uncertainty, in part because we don't know exactly what temperature to use for the estimate. The present result for $CO_2$-$Ar_9$ gives a similar estimate. We are not sure whether postulating such low frequency vibrations for $CO_2$-$Ar_n$ clusters is realistic, but it is difficult to think of another convincing explanation for the observed $CO_2$-$Ar_9$ and $CO_2$-$Ar_{17}$ sequences. It should be possible to investigate this question theoretically to decide whether our idea is feasible. A rigorous calculation is challenging because of multidimensionality (the many large amplitude intermolecular degrees of freedom). For example, $CO_2$-$Ar_9$ and $CO_2$-$Ar_{17}$ have 26 and 50 intermolecular modes, respectively, though many are degenerate which reduces the number of different modes. But exact calculations may not be necessary; approximate methods should be sufficient to establish the presence of one or more low frequency modes as suggested by our results.

## 7. Discussion and Conclusions

In the future, we hope to detect Kr and Xe clusters with n > 5, but their rotational structure will become increasingly difficult to resolve due to isotopic broadening and smaller rotational constants. Clusters with Ne are also of interest, and will require careful adjustment of the expansion gas mixtures. Even though we have reported the observation of $CO_2$-$Ar_n$ clusters up to n = 17, there are still a number of "missing" ones: n = 5, 7, 8, 13, 14, 16. Why are some clusters observed and not others? One factor is stability, since a more stable cluster is likely to be more abundant in the supersonic expansion. As discussed in Sec. 3, the chemical potential (Fig. 1) provides a measure of relative stability. Not surprisingly, more symmetric cluster structures (e.g., n = 9, 15, 17) tend to have chemical potentials of greater magnitude (minima in Fig. 1). Another factor affecting observability is whether the spectrum of a particular cluster can be rotationally resolved (at least in part) and recognized. Symmetric structure definitely makes a cluster spectrum more recognizable. Thus both abundance and recognizability favor symmetric cluster structures, and it is not surprising that our most secure assignments involve the symmetric structures n = 4 ($C_{2v}$), 9, 11, 15, and 17.

Many of the missing (unassigned) cluster sizes are almost certainly still present in our observed spectra, but remain unrecognized so far. The many otherwise unexplained lines mentioned in the text, and/or visible in Figs. 5, 7, 9, and 11, are all believed to be "real", in that they require the presence of $CO_2$ and Ar, Kr, or Xe. We also believe that they all involve clusters containing only one $CO_2$ molecule, though this is not absolutely sure (spectra of $(CO_2)_2$-Rg trimers were reported in Ref. 40).

The cluster rotational parameters reported here should be useful in searches for their pure rotational microwave transitions, and detection of such spectra would in turn help to better determine the cluster structures. Such searches would presumably start with smaller clusters (n = 3, 4) and with $CO_2$-$Ar_n$ since the Kr and Xe clusters are complicated by their multiple isotopes. Note that infrared observations have an advantage in that vibrational shifts move the spectra of different sized clusters into different spectral regions. In contrast, pure rotational spectra of different clusters tend to pile up on top of each other, though of course the much higher microwave spectral resolution serves to help separate them.

Despite longstanding theoretical interest, there have been few previous experimental results on $CO_2$-$Rg_n$ clusters with n > 2. The present infrared spectra (including Ref. 4) thus provide some of the first available data with which to test theoretical models. Having at least partial rotational resolution, the spectra yield rotational constants which provide direct structural information. Vibrational frequency shifts for the $CO_2$ $\nu_3$ mode are amenable to theoretical calculation and can help to interpret non-additive intermolecular force effects. The apparent observation of sequence bands arising from very low frequency thermally populated vibrational modes is intriguing and should also be amenable to theoretical verification

**Acknowledgements**

The financial support of the Natural Sciences and Engineering Research Council of Canada is gratefully acknowledged.

**References**

1   J. P. K. Doye, M. A. Miller, and D. J. Wales, Evolution of the potential energy surface with size for Lennard-Jones clusters, J. Chem. Phys. **111**, 8417-8428 (1999).

2   J. Tang and A. R. W. McKellar, High resolution infrared spectra of a carbon dioxide molecule solvated with helium atoms, J. Chem. Phys. **121**, 181-190 (2004).

3   A. R. W. McKellar, Infrared spectra of $CO_2$-doped ${}^4$He clusters, ${}^4$He$_N$-$CO_2$, with $N = 1 - 60$, J. Chem. Phys. **128**, 044308 (2008).

4   A. J. Barclay, A. R. W. McKellar, and N. Moazzen-Ahmadi, Observing the Completion of the First Solvation Shell of Carbon Dioxide in Argon from Rotationally Resolved Spectra, J. Phys. Chem. Letters **13**, 6311-6315 (2022).

5   J. M. Steed, T. A. Dixon, and W. Klemperer, Determination of the structure of $ArCO_2$ by radio frequency and microwave spectroscopy, J. Chem. Phys. **70**, 4095-4100 (1979); Erratum: **75**, 5977 (1981).

6   G. T. Fraser, A. S. Pine, and R. D. Suenram, Optothermal-infrared and pulsed-nozzle Fourier-transform microwave spectroscopy of rare gas–$CO_2$ complexes, J. Chem. Phys. **88**, 6157-6167 (1988).

7   H. Mäder, N. Heineking, W. Stahl, W. Jäger, and Y. Xu, Rotational spectrum of the isotopically substituted van der Waals complex Ar–$CO_2$ investigated with a molecular beam Fourier transform microwave spectrometer, J. Chem. Soc. Faraday Trans. **92**, 901-905 (1996).

8   R. W. Randall, M. A. Walsh, and B. J. Howard, Infrared absorption spectroscopy of rare-gas – $CO_2$ clusters produced in supersonic expansions, Faraday Discuss. Chem. Soc. **85**, 13-21 (1988).

9   S. W. Sharpe, R. Sheeks, C. Wittig, and R. A. Beaudet, Infrared absorption spectroscopy of $CO_2$-Ar complexes, Chem. Phys. Lett. **151**, 267-272 (1988).

10  S. W. Sharpe, D. Reifschneider, C. Wittig, and R. A. Beaudet, Infrared absorption spectroscopy of the $CO_2$–Ar complex in the 2376 cm$^{-1}$ combination band region: The intermolecular bend, J. Chem. Phys. **94**, 233-238 (1991).

11  Y. Ozaki, K. Horiai, T. Konno, and H. Uehara, Infrared absorption spectroscopy of Ar– ${}^{12}$C${}^{18}$O$_2$: change in the intramolecular potential upon complex formation, Chem. Phys. Lett. **335**, 188-194 (2001).

12  E. J. Bohac, M. D. Marshall, and R. E. Miller, The vibrational predissociation of Ar–$CO_2$ at the state-to-state level. I. Vibrational propensity rules, J. Chem. Phys. **97**, 4890-4900 (1992).

13  J. Thiévin, Y. Cadudal, R. Georges, A. A. Vigasin, Direct FTIR high resolution probe of small and medium size $Ar_n(CO_2)_m$ van der Waals complexes formed in a slit supersonic expansion, J. Mol. Spectrosc. **240**, 141-152 (2006).

14  T. A. Gartner, A. J. Barclay, A. R. W. McKellar, and N. Moazzen-Ahmadi, Symmetry breaking of the bending mode of $CO_2$ in the presence of Ar, Phys. Chem. Chem. Phys. **22**, 21488-21493 (2020).

15  M. Iida, Y. Ohshima, and Y. Endo, Induced dipole moments and intermolecular force fields of rare gas-$CO_2$ complexes studied by Fourier-transform microwave spectroscopy, J. Phys. Chem. **97**, 357-362 (1993).

16  T. Konno, S. Fukuda, and Y. Ozaki, Infrared spectroscopy of Kr–$^{12}C^{18}O_2$: Change in the $CO_2$ intramolecular potential by complex formation and isotope effect on the vibrationally averaged intermolecular geometry, Chem. Phys. Lett. **414**, 331-335 (2005).

17  T. Gartner, S. Ghebretnsae, A. R. W. McKellar, and N. Moazzen-Ahmadi, Spectra of $CO_2$ – Kr: intermolecular bend and symmetry breaking of the intramolecular $CO_2$ bend, Chemistry Select **2022**, e202202601 (2022).

18  A. J. Barclay, A. R. W. McKellar, C. M. Western, and N. Moazzen-Ahmadi, New infrared spectra of $CO_2$–Xe: modelling Xe isotope effects, intermolecular bend and stretch, and symmetry breaking of the $CO_2$ bend, Mol. Phys. **119**, e1919325 (2021).

19  Y. Xu, W. Jager, and M. C. L Gerry, Pulsed Molecular Beam Microwave Fourier Transform Spectroscopy of the van der Waals Trimer $Ar_2$-$CO_2$, J. Mol. Spectrosc. **157**, 132-140 (1993).

20  J. M. Sperhac, M. J. Weida, and D. J. Nesbitt, Infrared spectroscopy of $Ar_2CO_2$ trimer: Vibrationally averaged structures, solvent shifts, and three-body effects, J. Chem. Phys. **104**, 2202-2213 (1996).

21  A. J. Barclay, A. R. W. McKellar, and N. Moazzen-Ahmadi, Spectra of $CO_2$-$Rg_2$ and $CO_2$-Rg-He trimers (Rg = Ne, Ar, Kr, and Xe): intermolecular $CO_2$ rock, vibrational shifts and three-body effects, J. Chem. Phys. **157**, 204303 (2022).

22  J. Norooz Oliaee, B. Brockelbank, and N. Moazzen-Ahmadi, Use of quantum correlated twin beams for cancellation of power fluctuations in a continuous wave optical parametric oscillator for high-resolution spectroscopy in the rapid scan, The 25th Colloquium on High Resolution Molecular Spectroscopy, 20–25 August, Helsinki, Finland, 2017.

23 C. M. Western, PGOPHER, a program for simulating rotational structure version 8.0, 2014, University of Bristol Research Data Repository, doi:10.5523/bris.huflggvpcuc1zvliqed497r2

24 M. W. Severson, Quantum Monte Carlo simulations of $Ar_n$–$CO_2$ clusters, J. Chem. Phys. **109**, 1343-1351(1998).

25 M. Böyükata, E. Borges, J. C. Belchior, and J. P. Braga, Structures and energetics of $CO_2$-$Ar_n$ clusters (n = 1-21) based on a non-rigid model, Can. J. Chem. **85**, 47-55 (2007).

26 K. V. J. Jose and S. R. Gadre, An ab initio investigation on $(CO_2)_n$ and $CO_2(Ar)_m$ clusters: Geometries and IR spectra, J. Chem. Phys. **128**, 124310 (2008).

27 L. Wang and D. Xie, Simulated Annealing Study on Structures and Energetics of $CO_2$ in Argon Clusters, Chin. J. Chem. Phys. **24**, 620 (2011).

28 L. Wang and D. Xie, Finite temperature path integral Monte Carlo simulations of structural and dynamical properties of $Ar_N$−$CO_2$ clusters, J. Chem. Phys. **137**, 074308 (2012).

29 U. K. Deiters and R. J. Sadus, Two-body interatomic potentials for He, Ne, Ar, Kr, and Xe from ab initio data, J. Chem. Phys. **150**, 134504 (2019).

30 R. Chen, E. Jiao, H. Zhu, and D. Xie, A new ab initio potential energy surface and microwave and infrared spectra for the Ne–$CO_2$ complex, J. Chem. Phys. **133**, 104302 (2010).

31 Y. Cui, H. Ran, and D. Xie, A new potential energy surface and predicted infrared spectra of the Ar–$CO_2$ van der Waals complex, J. Chem. Phys. **130**, 224311 (2009).

32 R. Chen, H. Zhu, and D. Xie, Intermolecular potential energy surface, microwave and infrared spectra of the Kr–$CO_2$ complex from ab initio calculations, Chem. Phys. Lett. **511**, 229-234 (2011).

33 M. Chen and H. Zhu, Potential energy surface, microwave and infrared spectra of the Xe-$CO_2$ complex from ab initio calculations, J. Theo. Comp. Chem. **11**, 537-546 (2012).

34 Z. Wang, E. Feng, C. Zhang, and C. Sun, The potential energy surface and microwave spectra of the Xe–$CO_2$ complex, Chem. Phys. Lett. **619**, 14-17 (2015).

35 H. Li and R. J. Le Roy, Analytic three-dimensional 'MLR' potential energy surface for $CO_2$–He, and its predicted microwave and infrared spectra, Phys. Chem. Chem. Phys. **10**, 4128-4137 (2008).

36  W.H. Press, S.A. Reukolsky, W.T. Vetterling, and B.P. Flannery, Numerical Recipes in FORTRAN, Second Edition, Cambridge University Press (1992).

37  K. Mizuse, Y. Tobata, U. Sato, and Y. Ohshima, Rotational spectroscopy of the argon dimer by time-resolved Coulomb explosion imaging of rotational wave packets, Phys. Chem. Chem. Phys. **24**, 11014-11022 (2022).

38  P. E. LaRocque, R. H. Lipson, P. R. Herman, and B. P. Stoicheff, Vacuum ultraviolet laser spectroscopy. IV. Spectra of $Kr_2$ and constants of the ground and excited states, J. Chem. Phys. **84**, 6627-6641 (1986).

39  L. Fredin, B. Nelander, and G. Ribbegard, On the dimerization of carbon dioxide in nitrogen and argon matrices, J. Mol. Spectrosc. **53**, 410-416 (1974).

40  A.J. Barclay, A. R. W. McKellar, and N. Moazzen-Ahmadi, Infrared spectra of $(CO_2)_2$ - Rg trimers, Rg = Ne, Ar, Kr, and Xe, J. Mol. Spectrosc. **387**, 111673 (2022).